 \documentclass[aps,prb,reprint,superscriptaddress]{revtex4-1}
\usepackage{graphicx}
\usepackage[usenames, dvipsnames]{color}
\usepackage{gensymb}
\usepackage{ifthen}
\usepackage[version=3]{mhchem}
\usepackage{miller}
\usepackage{xspace}
\usepackage{textcomp}
\usepackage{nicefrac}

\usepackage{multirow}
\usepackage{longtable}
\usepackage{bm}

\usepackage[normalem]{ulem}

\newboolean{printSections} %Deklaration
\setboolean{printSections}{true} %Zuweisung

\newcommand{\Ca}{\ce{CaBaCo2Fe2O7}\xspace}
\newcommand{\ExE}{$\sqrt{3}\times\sqrt{3}$ }

\newcommand{\DM}{Dzyaloshinskii--Moriya }
\newcommand{\Jin}{J_\textrm{in}}
\newcommand{\Jout}{J_\textrm{out}}

\begin{document}
%\it
%\preprint{}

\title{Neutron diffraction study and theoretical analysis of the antiferromagnetic order and  \\ diffuse scattering  in the layered Kagome system \ce{CaBaCo2Fe2O7}}% with a chiral twist}


\author{J. D. Reim}
\email{johannes.reim@rwth-aachen.de}
\affiliation{Institute of Multidisciplinary Research for Advanced Materials,
Tohoku University, 2-1-1 Katahira, Sendai 980-8577, Japan}
\affiliation{J\"ulich Center for Neutron Science JCNS and Peter Gr\"unberg Institute PGI, JARA-FIT, Forschungszentrum J\"ulich, 
52425 J\"ulich, Germany}
%\thanks{}
\author{E. Ros\'en}
\affiliation{J\"ulich Center for Neutron Science JCNS and Peter Gr\"unberg Institute PGI, JARA-FIT, Forschungszentrum J\"ulich, 
52425 J\"ulich, Germany}
\author{O. Zaharko}
\affiliation{ Laboratory for Neutron Scattering and Imaging, Paul Scherrer Institute, 5232 Villigen, Switzerland }
\author{M. Mostovoy}
\affiliation{ Zernike Institute for Advanced 
Materials, University of Groningen, Nijenborgh 4, 9747 AG Groningen, Netherlands}
\author{J. Robert}
\affiliation{Institut NEEL, CNRS, 38042 Grenoble, France}
\author{M. Valldor}
\affiliation{Leibniz Institute for Solid State and Materials Research, 01069 Dresden, Germany}
%\author{O. Gourdon} \affiliation{Los Alamos National Laboratory, Los Alamos, USA}
\author{W. Schweika}
\email{werner.schweika@esss.se}
\affiliation{J\"ulich Center for Neutron Science JCNS and Peter Gr\"unberg Institute PGI, JARA-FIT, Forschungszentrum J\"ulich, 
52425 J\"ulich, Germany}
\affiliation{European Spallation Source ESS ERIC, Lund, Sweden}


\date{\today}

\begin{abstract}
The hexagonal swedenborgite, \Ca, is a chiral frustrated antiferromagnet, in which magnetic ions form alternating Kagome and triangular layers.
We observe a long range $\sqrt{3} \times \sqrt{3}$ antiferromagnetic order setting in below $T_N = 160$\,K by neutron diffraction on single crystals of \Ca.
Both magnetization and polarized neutron single crystal diffraction measurements show that close to $T_N$ spins lie predominantly in the $ab$-plane, while upon cooling the spin structure becomes increasingly canted due to \DM interactions.
The ordered structure can be described and refined within the magnetic space group $P31m^\prime$. 
Diffuse scattering between the magnetic peaks reveals that the spin order is partial.
Monte Carlo simulations based on a Heisenberg model with two nearest-neighbor exchange interactions show a similar diffuse scattering and  coexistence of the $\sqrt{3} \times \sqrt{3}$ order with disorder.
The coexistence can be explained by the freedom to vary spins without affecting the long range order, which gives rise to ground-state degeneracy.
Polarization analysis of the magnetic peaks indicates the presence of long-period cycloidal spin correlations resulting from the broken inversion symmetry of the lattice, in agreement with our symmetry analysis.

\end{abstract}

%\keywords{}

\maketitle

%%%%%%%%%%%%%%%%%%%%%%%%%%
%
% Introduction
%
%
\ifprintSections
\section{Introduction}
Frustrated magnetism gives rise to a wide range of exciting phenomena, such as complex non-collinear spin structures, nematic orders, spin liquid behavior and strong responses resulting from  large degeneracy of magnetic states.\cite{Ramirez1994,Moessner2006} 
Non-collinear spin orders, e.g. spirals and skyrmions, are also found in chiral magnets, where they are stabilized by  relativistic Dzyaloshinskii-Moriya interactions.\cite{Dzyaloshinskii1964,Moriya1960}
Recently synthesized Fe- and Co based compounds with the hexagonal swedenborgite structure\cite{Pauling1935,Valldor2002,Valldor2004a} are both magnetically frustrated and chiral. 
Magnetic transition metal ions in swedenborgites form alternating Kagome and triangular layers, which leads to geometric frustration.
The inversion symmetry is broken by the parallel alignment of oxygen tetrahedra coordinating the magnetic ions. 
Despite strong antiferromagnetic (AFM) exchange interactions between neighboring spins, most of these compounds show no long range magnetic order down to lowest temperatures and exhibit instead spin-liquid or spin-glass like behavior.\cite{Valldor2009}
Diffuse neutron scattering reveals  a variety of spin correlations in magnetically disordered swedenborgites: strongly one-dimensional correlations along the direction normal to the Kagome layers in YBaCo$_4$O$_7$ \cite{Manuel2009,Khalyavin2010} and  YBaFeCo$_3$O$_7$ \cite{Valldor2011} and  two- or three-dimensional correlation patterns in Y$_{0.5}$Ca$_{0.5}$BaCo$_4$O$_7$.   \cite{Schweika2007, Stewart2011}
The compound \Ca studied in this work is the only swedenborgite reported so far that shows a three-dimensional long range AFM order within the ideal hexagonal structure.\cite{Reim2014}
In the first approximation, swedenborgites can be described by a  Heisenberg model with only two exchange interactions between classical spins: $J_{\mathrm {in}}$  (between the nearest-neighbor spins in the Kagome plane) and $J_{\mathrm {out}}$ (between the neighboring Kagome and triangular sites).
For a suitable choice of the ratio $\tau=  J_{\mathrm {out}}/J_{\mathrm {in}}$, this model provides a good description of the observed diffuse scattering in YBaCo$_4$O$_7$\cite{Manuel2009} and reveals a variety of partially ordered and disordered ground states originating from the geometric frustration of the swedenborgite lattice structure.\cite{Khalyavin2010}
This behavior originates from the double tetrahedra building blocks formed by the magnetic ions.
The ground states of classical spins satisfy constraints (a ``sum rule''\cite{Manuel2009, Khalyavin2010}), which determine their degeneracy and spin ordering.
Recent numerical studies\cite{Buhrandt2014} showed that a large part of the phase diagram of this model is occupied by a spin-liquid state (see Fig.~\ref{phase_diagram}), in which the dimensionality of spin correlations depends on temperature and $\tau$. 
The phase diagram includes the coplanar phase characteristic of the two-dimensional Kagome lattice\cite{Reimers1993,Zhitomirsky2008} and the region of a nematic order selected by the entropic order-by-disorder mechanism as well as the magnetically ordered state stabilized above a critical value of $\tau$.

 \begin{figure}[h]
\begin{center}
\includegraphics[angle=0.0,  trim = 5mm 23.5mm 100mm 0mm, clip,width=6.2cm]
{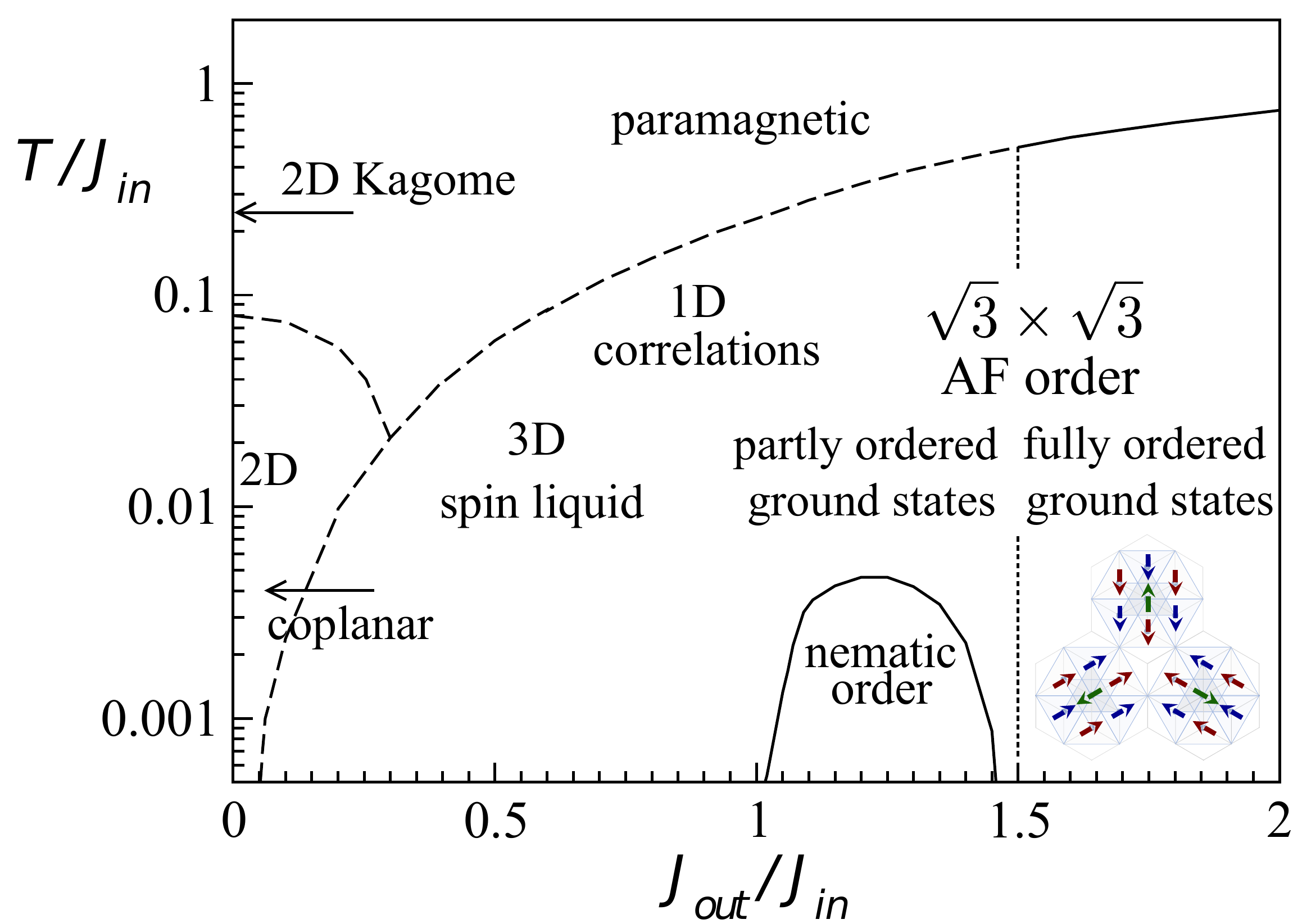}
\caption{\label{phase_diagram} Phase diagram  adapted from Ref.~[\onlinecite{Buhrandt2014}] with additional labels from Ref.~[\onlinecite{Khalyavin2010}].
For $\tau = J_{\mathrm {out}}/J_{\mathrm {in}} \ge 1.5$, the ground state is fully ordered in a $\sqrt 3 \times \sqrt 3$ AF structure.
For $\tau < 1.5$, a manifold of ground state degeneracy appears leading to a complex  and rich variety of spin correlations. Note the AF order persists into the proposed spin-liquid regime.}
\end{center}
\end{figure}

Here, we report the results of  neutron scattering experiments on high-quality large single crystals of \Ca.
We determined the  structure of magnetically ordered states of this compound  and studied the residual spin disorder by diffuse neutron scattering. 
We used rigorous polarization analysis of the Bragg and diffuse neutron scatterings to determine anisotropy of long-range magnetic orders and short-range spin correlations.
The same technique is used to probe the spin chirality vector
\begin{equation}
  \mathbf C =  [\mathbf S_{ \mathbf R}  \times \mathbf S _ { \mathbf {R}'}]
\end{equation} %\mathbf S_{ x} \times \mathbf S'_{ y}
defined on the bond connecting the canted spins $\mathbf S$ at the sites $\mathbf{R}$ and $\mathbf{R}'$.
The chirality originates from the \DM (DM) interactions which are allowed in the swedenborgite lattice as the tetrahedral oxygen coordination of magnetic sites breaks the inversion symmetry.
We performed Monte Carlo simulations based on the Heisenberg model focusing on the diffuse scattering observed experimentally and discuss the role of DM interactions.
The observation of diffuse scattering can be traced back to a degeneracy region found by analysing the energy of the ground state.
Considering the magnetic ordering, the structure of the intensity distribution indicates long periodic modulations to be energetically favored, whose nature was investigated using symmetry analysis.

\section{Experiment and Results}
%%%%%%%%%%%%%%%%%%%%%%%%%%
% Experimental methods
\ifprintSections
\subsection{Experimental Methods}
\fi

%%%%%%%%%%%%%%%%%%%%%%%%%%
The  single crystal specimen, 0.03\,ccm and 2\,ccm, for magnetization measurements and neutron diffraction experiments 
respectively,  were synthesized using the same method described previously in detail\cite{Reim2014}.

Magnetization measurements were performed with a  vibrating sample magnetometer option from Quantum Design. % \cite{Dynacool}.  
%a physical property measurement system (PPMS) named
% The ultra low field measurements have been performed at Tohoku University.
Both field cooled (FC) and zero field curves (ZFC) were measured. For the FC curve the field was applied at 350\,K then the 
sample was cooled down, while for the ZFC curve the field was applied at base temperature. Upon heating the magnetic moment was measured.

Single crystal neutron scattering experiments were performed on the cold neutron instruments DNS (Diffuse Neutron Scattering) 
at the Heinz Maier-Leibnitz Zentrum (Garching, Germany) and Morpheus at the Swiss Spallation Neutron Source SINQ, Paul Scherrer Institute (Villigen, Switzerland). 

%using polarized neutrons and polarization analysis (PA). 

Data for structure refinement were taken on Morpheus in the four circle geometry with 40' collimation at 4.7\,\AA\ using a single He-3 detector integrating intensities in rocking scans. 
In addition, to distinguish the magnetic scattering contributions near the Bragg peaks with high Q resolution, 
we performed polarization analysis %with both the guide fields and a field on the {$\omega$-circle} from permanent magnets in combination with multilayer polarizers for the incoming and final beam path and a Mezei-type flipper. 
using multilayer polarizers before and after the sample and a Mezei-type flipper in front of the sample. 
The polarization was maintained by permanent magnets supplying guide fields on the incoming and final beam paths and around the sample mounted on the Eulerian cradle.

Diffuse scattering was measured on DNS with a wavelength of 4.7\,\AA\ using XYZ polarization analysis for a wide angular detector coverage in the horizontal scattering plane.\cite{Schweika2001} 
Here, we measured all of the diagonal polarization channels for $x$, $y$, $z$-directions  as well as the spinflip, non-spinflip and  polarization reversal processes, which allows for a distinction of in-plane and out-of-plane magnetic scattering and in particular also for the determination of chiral magnetic scattering.

\subsection{XYZ-polarization analysis}
%It may be noted, that here $M_z$ is exclusively used to refer to said scattering contribution.
For XYZ-polarization analysis the neutron polarization $\mathbf P$  at the sample is rotated subsequently by applied magnetic fields into the orthogonal directions $x$, $y$, $z$.
The coordinate system follows the usual  convention of $ \hat {\mathbf e}_x || \mathbf Q$,  $\hat {\mathbf e}_y$ in the horizontal scattering plane and $\hat {\mathbf e}_z$ vertical. 
This also applies to the Fourier transforms $M_{\nu}({\mathbf Q}) $ of the magnetic moments with $\nu \in \{x, y, z\}$. 
Note that due to dipolar interaction only magnetic moments  perpendicular to the scattering vector $\mathbf Q$ contribute to magnetic scattering.
Hence, $\mathbf{M}_{\perp \mathbf Q} = (0, M_y, M_z) $, and  $M_x$ is zero.
$I_{\nu\nu}$ and $I_{\nu\overline{\nu}}$ are defined as the non-spinflip and spinflip scattering intensity, with the first and second indices  referring to the polarization direction of the incident  and scattered neutrons, respectively.
Combining different channels allows to separate the different magnetic scattering contributions\cite{Scharpf1993,Schweika2010}:
%Here, of particular interest are first, 
the total magnetic scattering $\left|\mathbf{M}_{\perp \mathbf Q}\right|^2 = \frac{1}{2}(I_{x\overline{x}} + I_{\overline{x}x})$, 
%second, 
the scattering from moments in the horizontal scattering plane $\left| M_y\right|^2 \approx \frac{1}{2}(I_{z\overline{z}} + I_{\overline{z}z})$,
%third, 
the scattering from moments perpendicular to this plane $\left|M_z\right|^2 \approx \frac{1}{2}(I_{zz} + I_{\overline{z}\overline{z}})$.
% and 
Alternatively, a simplified analysis (used for the four-circle measurements) is possible from the spin-flip and non-spinflip intensities with $\mathbf P \parallel \mathbf e_z$,
$\left|M_y\right|^2  \approx  I_{\overline{z}z}$ and $\left|M_z\right|^2 \approx I_{zz}  $, by neglecting weak contributions from nuclear coherent and incoherent scattering.

Finally, 
the chiral magnetic scattering is obtained by
\begin{eqnarray}\label{eq:ichiral}
 I_\textrm{chiral} = 
 \frac{1}{2}(I_{x\overline{x}} - I_{\overline{x}x})   & = & 
 i(\mathbf{M}^\dagger_{\perp\mathbf{Q}} \times \mathbf{M}_{\perp\mathbf{Q}}) \cdot \hat{\mathbf e}_x     \nonumber \\
 & \equiv & \mathbf{C}(\mathbf {Q}) \cdot \hat{\mathbf e}_x.
\end{eqnarray}
The measured wave vector dependent chirality 
% \begin{equation}\label{eq:chiralq}
$\mathbf{C}(\mathbf {Q})$  %=  i (\mathbf{M}^\dagger_{\perp\mathbf{Q}}   \times \mathbf{M}_{\perp\mathbf{Q}})
%\end{equation}
results from the sinus-Fourier transform of $M_y M_z$, the spin components perpendicular to $\mathbf{Q}$. 
Its antisymmetric property with respect to the propagation vector is sensitive to the specific chirality of the system and distinguishes the helical and cycloidal character of spiral spin structures. 
% 
% \begin{figure}[h]
% \includegraphics[angle=0.0, clip, width=7cm]
% {pictures/helix_cycloid_e_6.pdf}
% %\includegraphics[angle=0.0, %trim = 40mm 25mm 25mm 30mm, clip, width=7.3cm] {./pictures/Magnetization.pdf}%
% \caption{\label{fig:helixcycloid} \JR
% Helical and cycloidal spirals defined by a spiral wave vector $\mathbf{Q}_s$ and chirality vector $\mathbf C$.
% The color indicates the observable chiral magnetic scattering depending on the angle $\phi$ between $\mathbf{Q}_s$ and the scattering vector $\mathbf{Q}$ (red: positive chirality, blue: negative chirality, green: vanishing chirality).
% Thus, the resulting chiral magnetic scattering pattern is distinct for the type of the spiral.
% % As the chiral magnetic scattering is sensitive to the component of $\mathbf C$ parallel to the scattering vector $\mathbf Q$, it vanishes for cycloids if $\mathbf{Q}\parallel\mathbf{Q}_s$ .
% %in an antisymmetry (indicated by red and blue spheres) perpendicular to .
% }
% \end{figure}
% 
Amongst all chiral structures and their possible orientations,
the chiral scattering  $\mathbf C  \cdot \hat{\mathbf e}_{x}$ % in the $xy$-plane
%is sensitive to a helix  propagating parallel to $\mathbf Q$ with an antisymmetry along $\mathbf Q$, and similarly to a cycloid propagating parallel to $\hat{\mathbf e}_y$,
is sensitive to the components of a helix propagating parallel to $\mathbf Q$ and shows an antisymmetry along $\mathbf Q$.
Similarly, it is sensitive to the components of a cycloid propagating perpendicular to $\mathbf Q$, showing an antisymmetry along this direction. % (see Fig.~\ref{fig:helixcycloid})
%{\JR is sensitive to thes component of propagation vectors of helices parallel to $\mathbf Q$ and the component of propagation vectors of cycloids perpendicular to $\mathbf Q$.
%hence with an antisymmetry perpendicular to $\mathbf Q$.
In the case of the wide angular detector coverage of DNS, the more general approach for multi-detector systems is applied\cite{Schweika2010}.

\subsection{Magnetic properties}
The magnetic susceptibility of a  single crystal specimen measured in an external  field of 0.1\,T parallel and perpendicular to the $c$-axis is displayed in Fig.~\ref{fig:magnetization}.  
Below $T_\textrm{N}=160$\,K the magnetic susceptibility depends on the applied field direction. 
The kink observed for $H \perp c$ reflects the onset of an antiferromagnetic order and the stronger response to $H \parallel c$  indicates that the ordered spins are oriented predominantly in the $ab$-plane.
However, at low temperatures the magnetic susceptibiities for  $H \parallel c$  and $H \perp c$ gradually converge,  indicating some kind of spin reorientation. 
The induced moments per transition metal (TM) ion are rather small,  as can be expected for a system with strong AFM exchange interactions. 

\begin{figure}[h]
\includegraphics[angle=0.0, trim = 0mm 0mm 32mm 0mm, clip, width=5cm]
{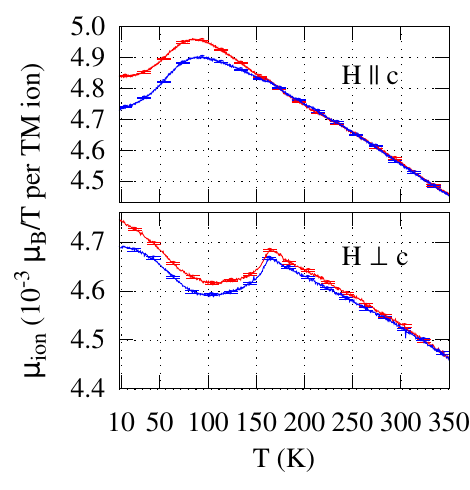}
\caption{\label{fig:magnetization} 
Temperature dependence of the zero-field cooled (blue line) and field cooled (red line) magnetization  induced by an external field of $H=0.1$\,T. Below 160\,K, the magnetic susceptibility of \Ca becomes anisotropic, being larger for $H \parallel c$ than for  $H \perp c$, which indicates the onset of an antiferromagnetic order with a preferrential orientation of the ordered spins  in the $ab$-plane.
}
\end{figure}

Remarkably, the zero-field and field cooled data split at low temperatures. 
Measurements of the field dependence of the magnetization reveal the presence of  a  weak ferromagnetic moment with a remanence of $\pm5\times 10^{-5}\,\mu_\textrm{B}$ per TM ion, which is essentially independent of the applied field direction and requires  fields of  $\sim 2$ T for closing the hysteresis loop at low temperatures.

\subsection{Magnetic structure determination}
For magnetic structure determination from neutron single crystal diffraction, we have collected two data sets well below the AFM ordering temperature,  $T_N=160$\,K, namely at 80 and 4\,K. 
This choice is  also motivated by  the magnetic susceptibility data showing the largest anisotropy at 80\,K and a spin reorientation at lower temperatures.
The data sets contain 116 magnetic peaks  in each case at the magnetic ordering  propagation vector $\mathbf q_c= (\nicefrac{1}{3},\nicefrac{1}{3},0)$ of the crystallographic reciprocal lattice units, {\it e.g.} the $K$-point of the {corresponding} Brillouin zone (BZ) boundary.  
Intensities were taken in rocking scans with appropriate background subtraction.
No attempt was made to determine the ferromagnetic contributions to the fundamental peaks of the crystallographic unit cell, because these contributions  are unmeasurably small according to the magnetization measurements. 

In the magnetic structure analysis, we considered  the   six different magnetic space groups proposed by the software MAXMAGN\cite{Perez-Mato2015} from the Bilbao crystallographic server for the larger unit cell 
  \((3a,3b,c)\) based on the position of magnetic ions and the propagation vector $\mathbf q_c$: \(P6_3c^\prime m^\prime\) (\#185.201), \(P6_3^\prime cm^\prime\) (\#185.200), \(P6_3^\prime c^\prime m\) (\#185.199), \(P6_3cm\) 
(\#185.197), \(P31m^\prime\) (\#157.55) and \(P31m\) (\#157.53). 
% 
%For magnetic structure determination from neutron single crystal diffraction, we have collected two data sets 
%well below the AF ordering appearing below  $T_N=160 K$, namely at 80 and 4 K. 
%The choice is  motivated also by  the magnetization data, suggesting the largest anisotropy at 80 K and a spin reorientation at lower temperatures. 
%The data sets contain 116 magnetic peaks  in each case, at the magnetic ordering 
%propagation vector $\mathbf q_c= (1/3,1/3,0)$ of the crystallographic reciprocal lattice units, {\it e.g.} the $K$-point of the {corresponding} Brillouin zone (BZ) boundary.  Intensities are taken in rocking scans integrating over any diffuse scattering or fine structures close to the magnetic Bragg peaks.
%No attempt was made to determine magnetic contributions on the fundamental peaks of the crystallographic unit cell, because the ferromagnetic contributions  are unmeasurably small according to the magnetization measurements. In fact,  we rather introduced additional fundamental peaks of zero magnetic intensity in accordance with the magnetization measurements.
%   
%
In the first four cases, the trigonal spins are restricted either to be parallel to the c-axis or to have zero moment for compatibility with the sixfold symmetry of the hexagonal space group.
The solutions with zero moment on the trigonal sites, \(P6_3^\prime c^\prime m\) and \(P6_3cm\),  can be discarded as inconsistent.   
The structural refinements within the remaining space groups were performed using  simulated annealing  within FullProf\cite{RodriguezCaravajal1993}.  
The refinements for all hexagonal space groups were significantly worse than for  $P31m^\prime$ and  $P31m$, which describe the measured data well.  
In order to reduce the number of free parameters, we assumed  that magnetic moments on the Kagome and  trigonal sites have the same magnitude respectively.
Furthermore, for two magnetic peaks only one of the symmetry equivalent ones could be measured, making these peaks prone to systematic errors.
Thus, these have been excluded from the refinements.
Note that a magnetic field applied perpendicular to the $c$ axis would lower the symmetry further, while an applied field or a ferromagnetic component parallel to the $c$ axis would break $P31m$ symmetry, it is compatible with \(P31m^\prime\) symmetry as all spins can cant out-of-plane. 
Following the order by disorder principle the latter solution with higher entropy is selected\cite{Villain1980}.
A comparison between the magnetic and fundamental peaks, using the crystallographic model of \Ca\cite{Reim2014}, allows for a refinement of the magnetic moments  on an absolute scale.
The results are shown in Table \ref{SASolution}.

\begin{table}[!ht]
\begin{ruledtabular}

\newcommand{\ch}[1]{\multicolumn{1}{c}{#1}}
\newcommand{\chb}[1]{\multicolumn{1}{c|}{#1}}
\newcommand{\ov}[1]{$^{\overline{#1}}$}

\begin{tabular}[c]{c|ccc|ccc}
& \multicolumn{3}{c|}{80\,K} & \multicolumn{3}{c}{4\,K} \\
site & $S$ [$\mu_B$]& $\Delta\phi$ [\textdegree]  &$\Delta\theta$ [\textdegree] & $S$ [$\mu_B$]& $\Delta\phi$ [\textdegree]  &$\Delta\theta$ [\textdegree] \\ \hline \rule{0pt}{2.5ex}
{tri},1    &2.73$^1$  & 180$^*$  &0.18\;  &2.63$^1$  &180$^*$ &-0.34\;	\\  \rule{0pt}{2.5ex}
{tri},2    &2.73$^1$  &180$^*$   & 0.01\;	&2.63$^1$  &180$^*$ & 0.01\;     \\  \rule{0pt}{2.5ex} 
{kag},1 &2.19$^2$  & 0.45\;    &11.19\;	&2.15$^2$  & 0.45\;  &15.84\;       \\     \rule{0pt}{2.5ex}
kag,2&2.19$^2$  & 0$^*$    &  -38.6\;   &2.15$^2$  & 0$^*$  &  -44.5\;  \\     \rule{0pt}{2.5ex} 
kag,3 &2.19$^2$  & 9.73\;    &-12.08\;    &2.15$^2$  & 11.89\; &-12.01 \;     \\ 	 \rule{0pt}{2.5ex}
kag,4 &2.19$^2$  & 0$^*$    &41.46\;     &2.15$^2$  & 0$^*$  &45.4\;      %\\	\hline
%R$_\text{Bragg}$ &\multicolumn{3}{c|}{17.4} & \multicolumn{3}{c}{22.8}
\end{tabular}
%{.~~~~~~~~~~~~80\,K}, \,R$_\text{Bragg}=${17.4}   {{~~~~~~4\,K}, \,R$_\text{Bragg}=$}{22.8}  

%\input{pictures/table_p31mp_modelD.tex}
\caption{Refinement results from simulated annealing applied to the reduced peak set for $P31m^\prime$. 
Only the independent spins are listed, others follow from the symmetry operations of the respective magnetic space group.
Parameters fixed by symmetry are marked with $^*$ and those linked by the model used in the refinement with same number $^n$. 
The spins on the triangular (tri) sites are on the special Wyckoff position $2a$ and the Kagome (kag) sites on $6c$. 
The offset of the azimuthal angle to the mirror plane through the center of the unit cell is noted as $\Delta \phi$.
$\Delta\theta$ denotes the deviation of the polar angle from the coplanar spin configuration (see inset of Fig.~\ref{phase_diagram}).}
\label{SASolution}
\end{ruledtabular}
\end{table}

\begin{figure}[htbp]
\begin{center}\includegraphics[trim = 20mm 65mm 20mm 60mm, clip, width=5.cm, clip, width=8.8cm]
{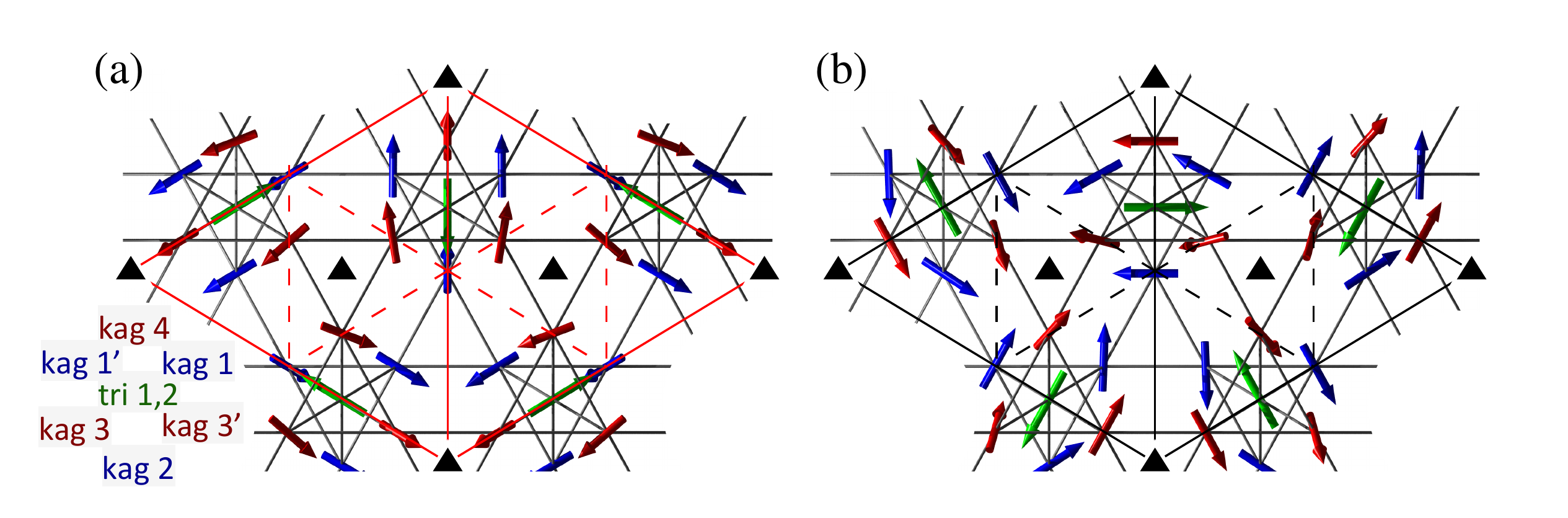}
\caption{
Magnetic structures with \(P31m^\prime\) symmetry (a)  and \(P31m\) symmetry (b), refined from a peak set of integrated intensities measured at 80\,K. 
We use the notations of Ref.~[\onlinecite{Litvin2008}] for magnetic symmetry operations symbols.
}
\label{Mag_structure}
\end{center}
\end{figure}

The magnetic structures with $P31m^\prime$ and $P31m$ symmetry obtained at 80\,K are displayed in Fig.~\ref{Mag_structure}.
Both are close to the 120$^\circ$ block spin ordering with the negative and positive vector chirality found for the Heisenberg model with two AFM exchange interactions,  $\Jin$ and $\Jout$, for $\tau =  \Jout/\Jin \geq3/2$  (see  the inset in Fig.~\ref{phase_diagram}).
The refined structures display the same antiparallel alignment between neighboring trigonal and Kagome spins. 
At the same time, where allowed by symmetry the especially the kagome spins cant in and out-of the $ab$-plane (see Table \ref{SASolution}) characteristic of a partial spin ordering in the Heisenberg model with  $\tau < 3/2$,  \cite{Khalyavin2010}.

\subsection{Magnetic scattering and  polarization analysis}
The measured integrated intensities of Bragg peaks only provide an insight into the average ordered part of the spin structure, and the relatively small ordered moment found in the structure determination indicates that the unordered part exists even at low temperatures.
To study the wave vector dependence of the magnetic scattering as well as the diffuse scattering related to disorder and short-range correlations in further detail, 
we performed a series of diffraction experiments, where we also applied polarization analysis to separate specific magnetic contributions. 
Part of these experiments were performed at the DNS instrument at MLZ providing a large overview of the reciprocal space. 
Using the XYZ-polarization analysis, the total magnetic scattering has been separated from the nuclear scattering and background\cite{Schweika2010}. 
\begin{figure}[h]
\begin{center}
\includegraphics[trim = 1mm 0mm 0mm 0mm, clip, width=5.5cm]{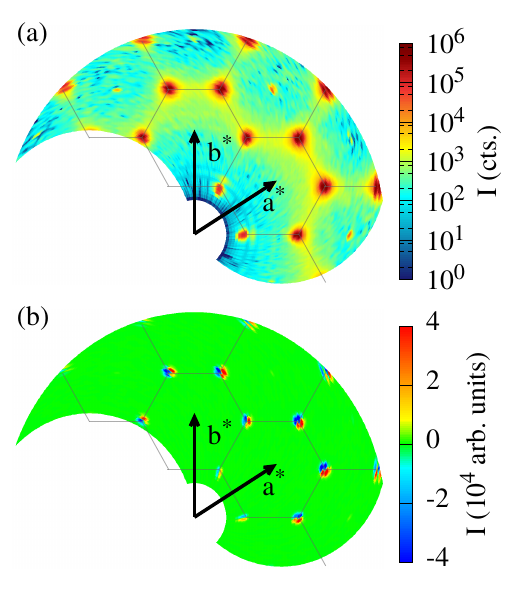}
\caption{
(a) Measured total magnetic scattering in the $(hk0)$-plane at $T=4\,$K. 
The magnetic intensity  $\left|\mathbf{M}_{\perp \mathbf Q}\right|^2$ has been separated by the XYZ polarization technique removing nuclear scattering and other background. 
The log-scale reveals significant diffuse scattering. 
(b) Chiral scattering component obtained from the difference in the spin-flip scattering with the polarization along and opposite to $\mathbf Q$ (red and blue spots). The antisymmetry perpendicular  to $\mathbf Q$ reveals the cycloidal character of spiral spin correlations.
\label{Diffuse_DNS}
}
\end{center}
\end{figure}
Figure~\ref{Diffuse_DNS}(a)  shows the magnetic scattering $\left|\mathbf{M}_{\perp \mathbf Q}\right|^2$ in the $(hk0)$-plane at 4\,K on a logarithmic scale to better reveal the diffuse scattering.
The magnetic peaks are found at the $K$-points of the BZ corresponding to a propagation vector $ \mathbf q_c= (\nicefrac{1}{3},\nicefrac{1}{3},0)$.
Additional measurements in $(hhl)$- and $(h0l)$-scattering planes do not show any further magnetic superstructure.
Significant diffuse scattering is seen near the magnetic Bragg peaks and along the Brillouin zone boundaries, 
indicating that the magnetic structure is only partially ordered and has a large  entropy even at 4 K. 
The chiral intensity related to $\langle \mathbf S \times \mathbf S' \rangle$, 
has been obtained by polarization reversal and is shown in Fig.~\ref{Diffuse_DNS}(b).
The antisymmetry of the chiral signal in $\mathbf Q$-space with positive and negative intensities near the $K-$points is shown here on a linear scale. 
While a helical spiral would give rise to an antisymmetric intensity along the scattering vector $\mathbf Q$, the observed antisymmetry perpendicular to $\mathbf Q$ is characteristic of a cycloidal spiral modulation in the $(hk0)$-plane.

The intensity distribution at the $(\nicefrac{4}{3},\,\nicefrac{1}{3},\,0)$ position and its polarization dependence was studied in further detail, as shown in Fig.~\ref{Morpheus_peak} on a linear scale. 
Compared to the nuclear peaks, the magnetic peaks exhibit a clearly richer structure.
In addition to the total magnetic scattering [Fig.~\ref{Morpheus_peak}(a)], we show the contributions of the spin components parallel and perpendicular to the $ab$-plane [Figs.~\ref{Morpheus_peak}(b) and (c), respectively]. 
\begin{figure}[htbp]
\begin{center}
\includegraphics [width=7cm] {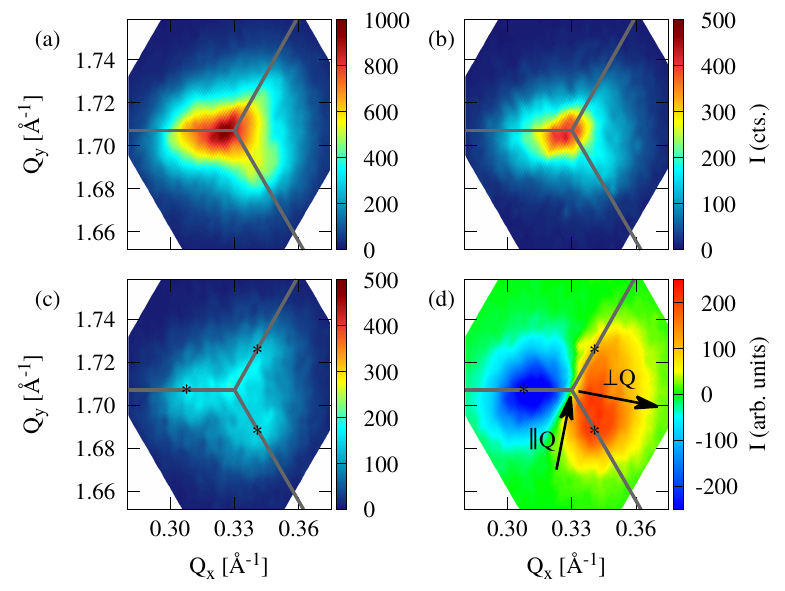}
\caption{The magnetic scattering at {$T=4 \, \mathrm K$} near the $(\nicefrac{4}{3},\,\nicefrac{1}{3},\,0)$ position with fine structure extending along the BZ boundaries: 
(a)  total scattering, 
(b) contribution from spin components in $ab$-plane, 
(c) contribution from spin components in $c$-direction, 
(d) chiral signal reveals a purely cycloidal character exhibiting an antisymmetry perpendicular to $\mathbf{Q}$.
Possible precursor of a long-periodic order are marked with *.
}
\label{Morpheus_peak}
\end{center}
\end{figure}
Apparently, the in-plane spin components, $|M_y|^2$, largely contribute to the peak maxima with $ \mathbf q_c$. 
The signal from the out-of-plane spin components, $|M_z|^2$, close to the $K$-point is weaker and more evenly distributed.
It can be seen as a precursor of a three-fold peak splitting due to the formation of long-period modulated structures. 
According to Eq.~\ref{eq:ichiral}, the chiral intensity describes  the correlation between the in-plane and out-of-plane spin components $M_yM_z$.
The pattern in Fig.~\ref{Morpheus_peak}(d) shows with high resolution the characteristic antisymmetry from cycloids  perpendicular to $\mathbf Q$.  % {\JR (see Fig.~\ref{fig:helixcycloid})}
The maxima positions in the chiral scattering coincide with two of the precursors, however, the propagation vector of the third one deviates slightly from $ \mathbf Q$ leading to an asymmetric yet weak chiral scattering.
Recent experiments actually revealed the formation of a complex long periodic order at low temperatures, which will be discussed separately\cite{Reim2018}. 

\begin{figure}
\includegraphics[angle=0.0,width=6cm]{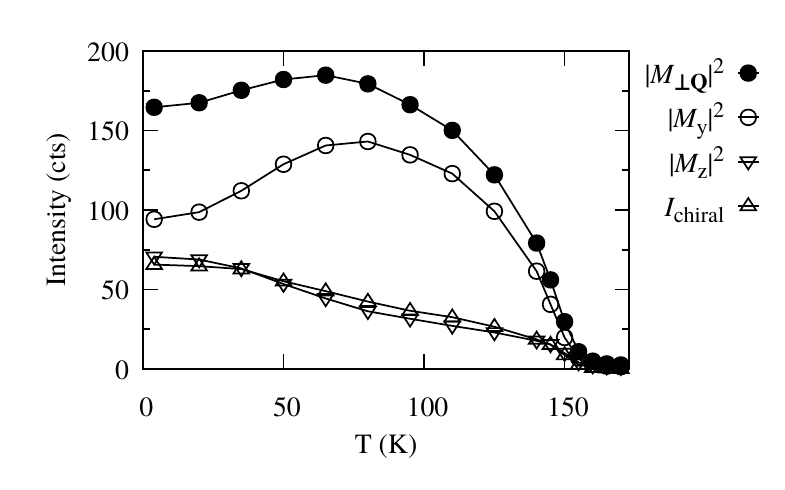}  
\caption{ \label{DNS-Tdep} 
Temperature dependence of the scalar and chiral magnetic scattering contributions near the peak $(\nicefrac{4}{3},\,\nicefrac{1}{3},\,0)$. 
}
\end{figure}

The DNS instrument was also used to measure temperature dependence of various magnetic contributions (see Fig.~\ref{DNS-Tdep}).
The temperature dependence of the integrated total magnetic  intensity  of the peak $(\nicefrac{4}{3},\,\nicefrac{1}{3},\,0)$  as well as  the contribution of the in-plane spin components seem to follow a typical continuous phase transition. 
In contrast, $\left|M_z\right|^2$ exhibits  an unusual temperature dependence, which closely  follows the magnitude of the chiral signal obtained by integrating  the absolute values of the antisymmetric  modulation. 
Although the chiral signal appears simultaneously with the magnetic order at $T_\textrm{N}$, it grows much slower below $T_\textrm{N}$ than the total peak intensity, which suggests that the chiral order has a larger critical exponent than the magnetic order parameter.
The out-of-plane spin contribution to the scattering increases with decreasing temperature (below 80\,K) and is accompanied by a decrease of the in-plane contribution, so that at lowest temperatures the two contributions become almost equal. 
This result is in agreement with the spin re-orientation  deduced from the temperature dependence of the magnetic susceptibility (see Fig.~\ref{fig:magnetization}).
The simultaneous increase of the chiral signal shows, however, that the low-temperature spin structure is not just more isotropic but that it is more complex than the high-temperature spin structure.

\section{Theoretical analysis and modeling} 
\subsection{Exchange interactions and Heisenberg model}
The  magnetic sublattice of \Ca is built up by corner sharing CoO$_4$ and FeO$_4$ tetrahedra forming alternating Kagome and triangular layers (see Fig.~\ref{exchange}).
The oxygen ions impose a tetrahedral crystal field splitting on transition metal sites, with $e_g$ levels being lower in energy than the $t_{2g}$ levels. 
The high-spin  Co$^{2+}$ and Fe$^{3+}$ ions have the electronic configuration $e_g^4t_{2g}^3$ and $e_g^2t_{2g}^3$, respectively, as depicted in Fig.~\ref{SE}(a). 

The Heisenberg exchange interactions between neighboring transition metal ions are expected to be antiferromagnetic \cite{Khalyavin2010,Chatterjee2011}. 
Moreover, Co-Co, Co-Fe and Fe-Fe interactions are expected to be of comparable strength, since both Fe and Co have three $t_{2g}$ electrons with parallel spins, which give the largest contribution to the superexchange \cite{Goodenough1963,Anderson1963,Khomskii2014}.  
The relevant orbitals are shown in Fig.~\ref{SE}(b), where we use a pseudocubic frame with the [111]-direction along the $c$-axis. 
The bond lengths  obtained  from single crystal neutron diffraction show deviations from the ideal tetrahedral stacking\cite{Reim2014}.
In our spin model of \Ca we neglect these deviations and consider only one in-plane and one out-of-plane nearest-neighbor exchange interaction constants,  $J_{\mathrm{ in}}$ and $J_{\mathrm{out}}$, respectively.

\begin{figure}[htbp]
\begin{center}
\includegraphics[angle=0.0,  trim = 155mm 22mm 30mm 112mm, clip, width=6.5cm]
{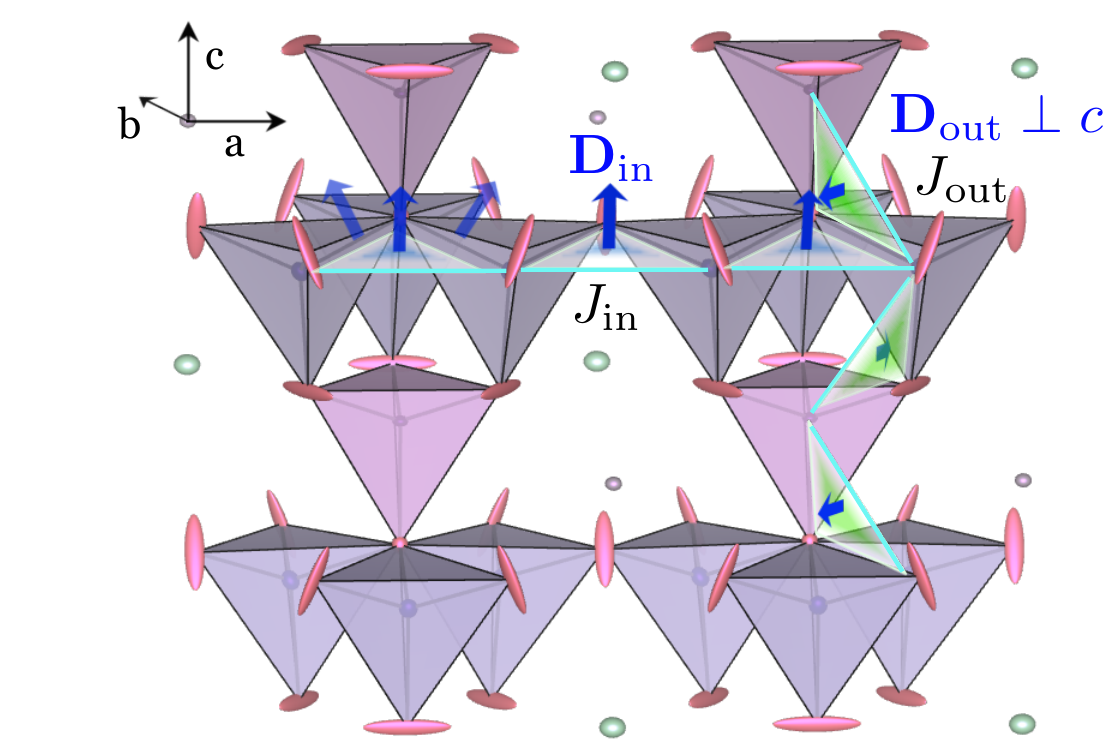}
\caption{Sublattice of magnetic ions located in the O-tetrahedra, exchange interactions (cyan lines) and Dzyaloshinskii-Moriya interactions (blue vectors).
 $\Jin$  denotes the nearest-neighbor exchange interaction in the Kagome layers and $\Jout$  is  the interaction between the  nearest-neighbor spins in the Kagome and the triangular layers.
 The triangles show schematically the superexchange path between the  magnetic ions via the O-sites. 
 The blue arrows indicate the directions of the Dzyaloshinskii-Moriya vectors, ${\bm D}_{\mathrm{in(out)}}$.
}
\label{exchange}
\end{center}
\end{figure}

\begin{figure}[htbp]
\begin{center}
\includegraphics[angle=0.0,trim = 00mm 91.5mm 221.5mm 00mm, clip, width=6cm] 
{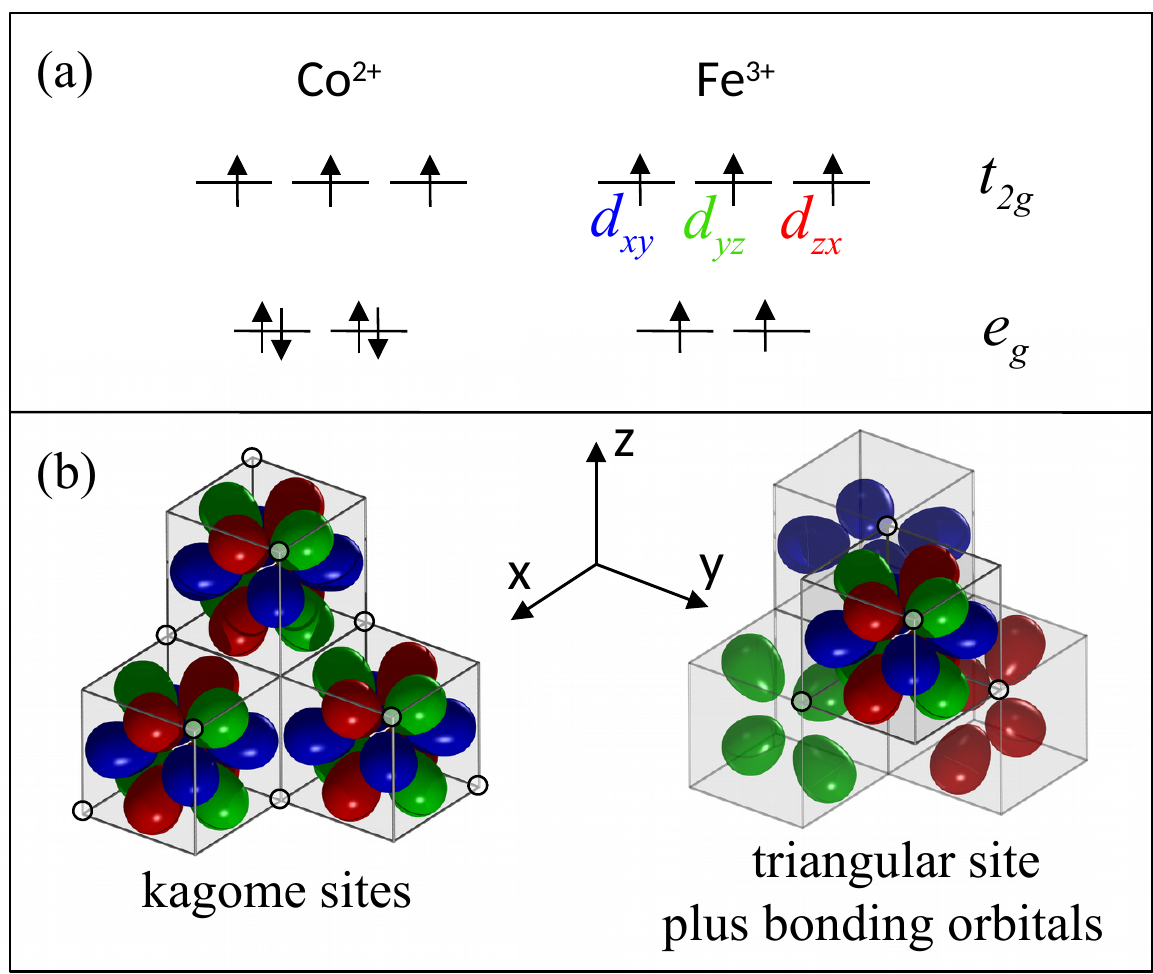}
\caption{
Super-exchange between half-filled $t_{2g}$-shells of  Co- and Fe-ions. 
(a) Crystal-field splitting and electronic configurations of tetrahedrally coordinated Co$^{2+}$  and Fe$^{3+}$ ions.  
(b) $t_{2g}$ orbitals on three neighboring Kagome sites ({\em left}),  which dominate the superexchange due to the relatively large overlap with the $p$-orbitals of oxygen ions at the edges of the tetrahedra. 
The same type of exchange takes place between the ions on Kagome and triangular sites ({\em right}).
}
\label{SE}
\end{center}
\end{figure}
While it is clear that additional interactions are needed to account for the observed magnetic anisotropy and chiral correlations, we first discuss our observations within a minimal Heisenberg model\cite{Khalyavin2010}:  
\begin{equation}
\label{HeisenbergModel}
H = - \Jin {\sum_{<i,j>}}' {\bf S}_i \cdot {\bf S}_j -  \Jout 
{\sum_{<i,j>}}''  {\bf S}_i \cdot {\bf S}_j
\end{equation}
with $\Jin,\Jout<0$ and $S_{i} = 1$. The sum in the first term goes over pairs of nearest-neighbor spins in the Kagome layers and the second sum is  over the bonds connecting nearest-neighbor triangular and Kagome sites. 

\subsection{Frustration parameter $\tau$}
We first consider a triangular bipyramid made of a Kagome triangle with the spins ${\bf S}_1$, ${\bf S}_2$ and ${\bf S}_3$, (red triangle in Fig.~\ref{fig:sites}) capped with trigonal sites above and below with the spins ${\bf S}_4$ and  ${\bf S}_5$. 
In the minimal energy state the spins on the trigonal sites are antiparallel to the total spin of the Kagome triangle, 
\begin{equation}
 {\bf S}_4 = {\bf S}_5 = - \frac{{\bf S}_{\triangle}}{S_{\triangle}},\label{eq:S4S5}
\end{equation}
where 
\begin{equation}
\label{eq:Striangle}
{\bf S}_{\triangle} = {\bf S}_1 + {\bf S}_2 + {\bf S}_3 
\end{equation}
and $S_{\triangle} = |{\bf S}_{\triangle}|$. Minimization of the  bipyramid energy, $\frac{1}{2} \Jin (S_{\triangle}^2 - 3)  + 2 \Jout S_{\triangle}$, with respect to  $S_{\triangle}$, gives\cite{Khalyavin2010}
\begin{equation}\label{eq:striag}
S_{\triangle} = \left\{
\begin{array}{cc}
3, & \tau \ge 3/2;\\
2\tau, & \tau < 3/2;
\end{array}
\right.
\end{equation}
where $\tau = \Jout / \Jin > 0$. For $\tau \ge 3/2$, ${\bf S}_1 =  {\bf S}_2 = {\bf S}_3$, while for $\tau < 3/2$ there are many ways to add these three spins into the total spin $2\tau$. 

Next we consider antiferromagnetic exchange interactions between the bipyramids. 
The bonds in the Kagome layers connecting three neighboring bipyramids form triangles (green triangle in Fig.~\ref{fig:sites}). 
The exchange energy is minimal, if the spins in all such interpyramidal triangles are added to 0.
For $\tau \ge 3/2$, this is achieved in the 120$^{\circ}$-state formed by the total spins of the bipyramids (see insert in Fig.~\ref{phase_diagram}). 
This ground state is unique up to the global rotation of all spins and the change of sign of the vector spin chirality. 
For $\tau < 3/2$, the freedom in adding the spins in the Kagome triangles into the total spin $2\tau$ gives rise to many ways in which the spins in interpyramidal triangles add to 0, resulting in degenerate classical ground states. 

\subsection{Spiral ground states}
%%%%
We consider a particular class of ground states -- the spiral ground states, constructed in the following way. We chose 8 spins
in one unit cell, which contains two spin triangles in neighboring Kagome layers and two spins on trigonal sites between them. 
Spins in other unit cells are then obtained from the 8 spins  by rotation  around an axis, ${\bf n}$, through the angle $\varphi = {\bf Q}_s\cdot {\bf R}$, where ${\bf Q}$ is the spiral  wave vector parallel to the $ab$ plane and ${\bf R}$ is the distance between the cells. 
Although the spiral ground states are a small part of all classical ground states, they have a massive   degeneracy: for $\tau < 3/2$, their energy is independent of the spiral wave vector ${\bf Q}_s$, for all  ${\bf Q}_s$ from a region $G_\tau$ of the reciprocal space.

Consider the interpyramidal triangle formed by the spins
${\bf S}_1$, ${\bf S}_{2-b}$ and ${\bf S}_{3+a}$ (green triangle in Fig.~\ref{fig:sites}), such that ${\bf S}_1+{\bf S}_{2-b}+{\bf S}_{3+a}=0$. In the spiral state ${\bf S}_{3+a} = \hat{R}_{\bf n}(\varphi_a){\bf S}_3$, where  $\hat{R}_{\bf n}(\varphi)$ is the operator of spin rotation around $\bf n$ through the angle $\varphi$ and $\varphi_a = 2\pi Q_{a^\ast}$, because the spin ${\bf S}_{3+a}$ is separated from  ${\bf S}_{3}$ by the vector ${\bf a}$. Similarly, ${\bf S}_{2-b} = \hat{R}_{\bf n}(-\varphi_b){\bf S}_2$, because ${\bf S}_{2-b}$ is separated from ${\bf S}_{2}$ by the vector $- {\bf b}$. We then have,
\begin{equation}
{\bf S}_1 + \hat{R}_{\bf n}(-\varphi_b){\bf S}_2 +  \hat{R}_{\bf n}(\varphi_a){\bf S}_3 = 0.
\label{eq:inter}
\end{equation} 
By construction, Eq.~(\ref{eq:inter}) guarantees that the total spin of {\em all}  interpyramidal triangles is 0. This equation  should hold together with $S_\triangle = \left | {\bf S}_1 + {\bf S}_2 + {\bf S}_3 \right | = 2\tau$ in order to minimize the total exchange energy Eq.~(\ref{HeisenbergModel}). 
\begin{figure}[htbp]
\includegraphics[angle=0.0,trim = 0mm 110mm 0mm 0mm, clip, width=7cm]{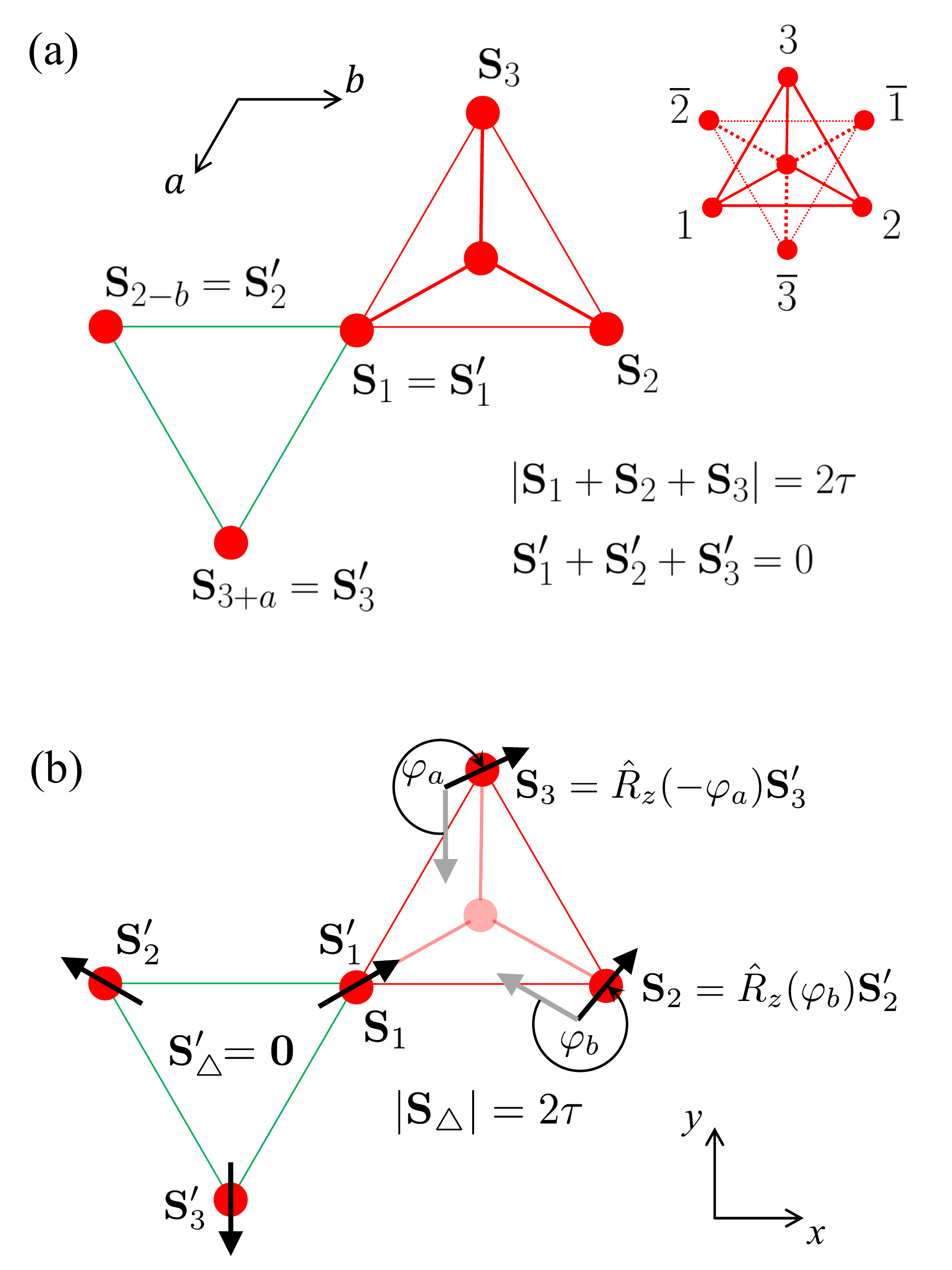}
\caption{ \label{fig:sites} 
(a) The spins  ${\bf S}_1$, ${\bf S}_2$ and ${\bf S_3}$ in the bipyramide triangle ($|{\bf S}_1 + {\bf S}_2 + {\bf S}_3| = 2 \tau$) and the spins in the bipyramidal triangles, ${\bf S}'_1 = {\bf S}_1$,  ${\bf S}'_2 = {\bf S}_{2-b}$ and ${\bf S}'_3 = {\bf S}_{3+a}$ forming the 120$^{\circ}$-state (${\bf S}'_1 + {\bf S}'_2 + {\bf S}'_3 = 0$). (b) In the spiral ground states spins ${\bf S}_2$ and ${\bf S}_3$ are obtained by rotating the spins ${\bf S}'_2$ and ${\bf S}'_3$ through the angles $\varphi_b$ and $-\varphi_a$, respectively.
 }
\end{figure}

It is convenient to introduce ${\bf S}'_1 = {\bf S}_1$, ${\bf S}'_2 = \hat{R}_{\bf n}(-\varphi_b){\bf S}_2$ and ${\bf S}'_3= \hat{R}_{\bf n}(\varphi_a){\bf S}_3$ [see Fig.~\ref{fig:sites}(a)], so that  Eq.~(\ref{eq:inter}) becomes 
\begin{equation}
{\bf S}'_1 + {\bf S}'_2 + {\bf S}'_3 = 0,
\label{eq:triangleprime1}
\end{equation}
and 
\begin{equation}
\left| {\bf S}'_1 + \hat{R}_{\bf n}(\varphi_b){\bf S}'_2 +  \hat{R}_{\bf n}(-\varphi_a){\bf S}'_3 \right| = 2 \tau.
\label{eq:triangleprime2}
\end{equation}
If Eqs.~(\ref{eq:triangleprime1}) and (\ref{eq:triangleprime2}) are satisfied simultaneously, then both the intrablock and interblock exchange energies  are minimized. Substituting ${\bf S}'_1 = - {\bf S}'_2 - {\bf S}'_3$ into Eq.~(\ref{eq:triangleprime2}), we find that the spin rotation axis of the spiral state is orthogonal to $ {\bf S}_\triangle$: ${\bf n} \perp {\bf S}_\triangle$.  In view of the rotational invariance of the Heisenberg model (\ref{HeisenbergModel}), we can assume that ${\bf n}$ is parallel to the $z$-axis.

We first assume that the three spins, ${\bf S}'_1$, ${\bf S}'_2$ and ${\bf S}'_3$ lie in the $xy$-plane and that the spin ${\bf S}'_1$ [see Fig.~\ref{fig:sites}(b)] is parallel to the $x$ axis. From Eq.~(\ref{eq:triangleprime2}) we then obtain for $\delta \varphi_{a,b} = \varphi_{a,b} - \frac{2\pi}{3} = 2\pi \left(Q_{a^{\ast},b^{\ast}} - \frac{1}{3}\right)$:
\begin{equation}
\cos \delta \varphi_a + \cos \delta \varphi_b + \cos (\delta \varphi_a + \delta \varphi_b) = \frac{4\tau^2 - 3}{2}.
\end{equation}
The solution of this equation is the contour $C_0$ in the reciprocal space, which lies in the $a^{\ast} b^{\ast}$-plane and has of a nearly circular shape with the center at the K-point ${\bf q}_K = (\nicefrac{1}{3},\,\nicefrac{1}{3},\,0)$ (white contour in Fig.~\ref{fig:degeneracy}).  

\begin{figure}
\includegraphics[angle=0.0, width=4cm]{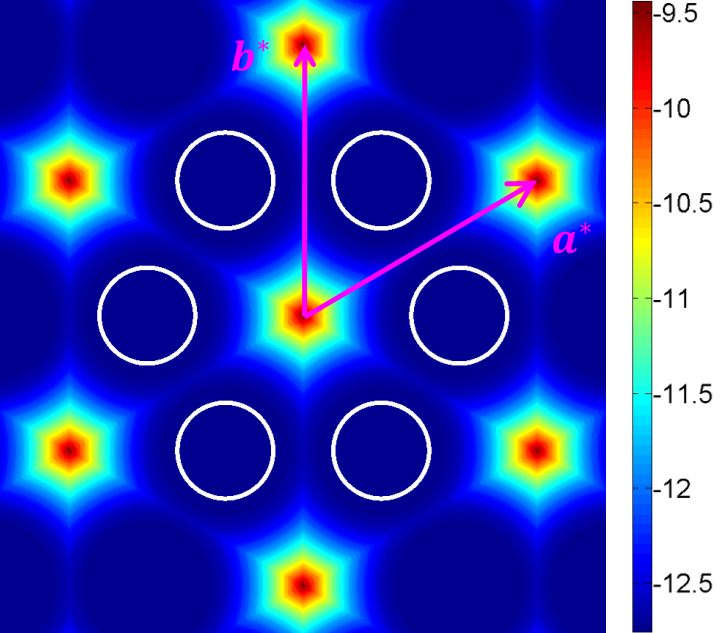}
\caption{
\label{fig:degeneracy} False color plot of the spiral state energy  in units of $\Jin$ per unit cell, for $\tau = 1.3$, as a function of the spiral wave vector, ${\bf Q}$. 
White circles mark the boundaries of the degeneracy regions, $G_\tau$, with the centers at the $K$-point $(\nicefrac{1}{3},\,\nicefrac{1}{3},\,0)$ and other symmetry-related points. 
Inside $G_\tau$ the energy is independent of the spiral wave vector and the energy landscape is perfectly flat.
}
\end{figure}

More ground states are found for  ${\bf S}'_2$ and ${\bf S}'_3$  obtained from those shown in Fig.~\ref{fig:sites}(b)  by rotation  around the $x$ axis through an angle $\psi$. 
Then the solution of Eq.~(\ref{eq:triangleprime2}) is a contour $C_\psi$ that lies in the region $G_\tau$ bounded by $C_0$. 
As $\psi$ continuously increases from 0 to its maximal possible value, $\psi_{\rm max} = \arccos\left(2 \sqrt{\tau(\tau  - 1)/3}\right)$, the solutions of Eq.~(\ref{eq:triangleprime2}) fill the whole $G_\tau$ (for $\psi =\psi_{\rm max}$, the contour $C_\psi$ shrinks into a single point).  
All thus obtained spiral states have the  minimal exchange energy for a given $S_\triangle  = 2 \tau$. The maximal value of $\psi$ only exists for $\tau > 1$ or $S_\triangle > 2$. 
For $\tau < 1$ the degeneracy regions around neighboring $K$-points overlap.

Moreover, for each ${\bf Q}_s$ inside the two-dimensional manifold  of spiral wave vectors, $G_\tau$, there exist many different spiral states with the same energy. 
The degeneracy results from the freedom to choose ${\bf S}'_1$, ${\bf S}'_2$ and ${\bf S}'_3$, which can be obtained from those  shown in Fig.~\ref{fig:sites}(b) by an arbitrary rotation parametrized by three Euler angles. 
They give rise to different sets,  $({\bf S}_1, {\bf S}_2, {\bf S}_3),$ for given $\varphi_a, \varphi_b$ and $S_{\triangle}$, which are, in general, inequivalent, i.e. cannot be obtained from each other by a global rotation of all spins. 
This leads to a massive ground state degeneracy for $\tau < 3/2$. 

Figure~\ref{fig:degeneracy} shows the false color energy plot as a function of the spiral wave vector, for $\tau = 1.3$, obtained by numerical minimization of energy with respect to the 8 spins from   
the magnetic building block of the swedenborgite spin lattice, for a given ${\bf Q}_s$. 
Inside the degeneracy regions, $G_\tau$, the boundaries of which are marked by white contours, the energy landscape is perfectly flat, i.e. independent of  ${\bf Q}_s$. 
We note here a simple relation between spins from neighboring Kagome triangles in the spiral ground states. 
We denote by ${\bf S}_{\bar{1}}$, ${\bf S}_{\bar{2}}$ and ${\bf S}_{\bar{3}}$ spins on the sites obtained by the $2_1$ symmetry operation  from the sites 1, 2 and 3, respectively [see Fig.~\ref{fig:sites}(a)]. 
The relation between the spins in two triangles of the same magnetic block in the minimal-energy spiral state is: ${\bf S}_{\bar{i}} = 2 \hat{{\bf S}}_\triangle (\hat{{\bf S}}_\triangle \cdot {\bf S}_i)- {\bf S}_{i}$ $(i = 1,2,3)$, where $\hat{{\bf S}}_\triangle = {\bf S}_\triangle / S_\triangle$ (the total spins of the two triangles are equal).
 
Although spin spirals are a subclass of all degenerate ground states of the model Eq.~(\ref{HeisenbergModel}), they are  selected by additional interactions present in \Ca, such as DM interactions. 
As discussed in below, the lack of inversion symmetry in the crystal lattice of the swedenborgite leads to Lifshitz invariants originating from the DM interactions, which favor incommensurate magnetic states. 
These interactions lift the degeneracy by selecting the spiral wave vector ${\bf Q}_s$. 

%%%%%%%%%%%%%%%%%%%%%%%%%%
% Modeling magnetic order from DNS
\ifprintSections
\subsection{Monte Carlo simulation of diffuse scattering}
\fi
%%%%%%%%%%%%%%%%%%%%%%%%%%

For a closer comparison to the observed diffuse scattering, we performed Monte Carlo simulations based on the nearest-neighbor Heisenberg model for varying ratios $\tau$ of in- and out-of plane interactions.
In this calculation, the spin lattice has been cooled slowly to low temperatures, $T/\Jin=0.01$, and averaged over 500 independent spin structures. 
A model  size $N= 8L^3$ with $L=24$ was chosen for the simulations in Fig.~\ref{MCS-order}.

Figure~\ref{MCS-order} shows the magnetic structure factor obtained by Monte Carlo simulations of the Heisenberg model Eq.~(\ref{HeisenbergModel}) for three ratios of the  in- and out-of plane interactions, $\tau$. 
For $\tau \ge 1.5$,  3D long-range antiferromagnetic order of $\sqrt 3 \times \sqrt 3$ type and vanishing diffuse scattering is found as expected \cite{Khalyavin2010, Buhrandt2014}.
%In general the structure factor for ratios $1.0\leq \tau < 1.5$ reveals prominent peaks at the K-points of the BZ, Bragg-like related to long-range antiferromagnetic order as well as diffuse scattering.
For $\tau < 1.5$, the simulated structure factor shows diffuse intensity, reflecting the ground state degeneracy discussed above, which is getting stronger as $\tau$ decreases.
However, prominent Bragg-like peaks at the K-points do not disappear for  $\tau < 1.5$, in apparent contradiction with the phase diagram of this model obtained earlier \cite{Buhrandt2014}. 
We looked for possible finite size effects at $\tau =1.3$, by comparing the results for the $L=12$ and 24 lattices and the system still appears to be antiferromagnetically ordered, while the calculated S({\bf Q}) agrees with the earlier calculation\cite{Buhrandt2014} for $L=9$.
We conclude that the AFM phase field extends further into the region $\tau <1.5$, albeit including residual disorder and entropy as seen by diffuse scattering.
At $\tau=1.1$, the diffuse scattering is more enhanced and compares well with the observed diffuse scattering.
Here, periodic boundary conditions and finite system size show weak effects on the reminiscent Bragg-like intensities, 
while only negligible size effects are seen for the diffuse scattering in Figs.~\ref{MCS-order} and \ref{Fig1-magord}.
The coexistence of Bragg peaks with strong diffuse scattering is consistent with our experimental observations.
Figure~\ref{Fig1-magord} shows the comparison of the observed diffuse scattering intensity being corrected for the magnetic form factor (see Panel a) with the  Monte Carlo simulations for  $\tau=1.1$ (see Panel b), being rescaled and both interpolated to a similar finer grid in the reciprocal space.  

\begin{figure}
\includegraphics[angle=0.0, trim = 0mm 0mm 0mm 0mm, clip, width=7.5cm]{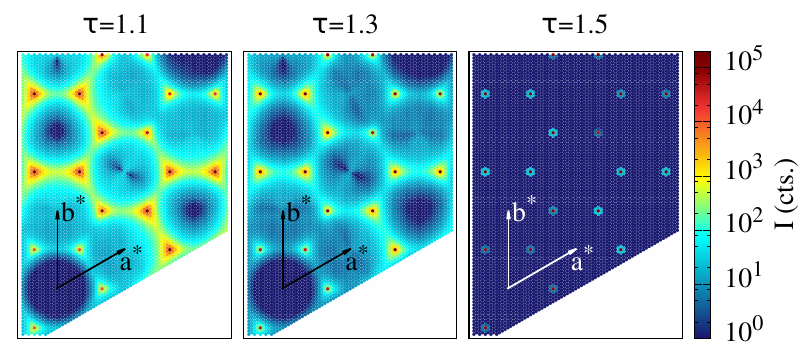}
\caption{\label{MCS-order} 
Magnetic structure factor in the ($hk0$) plane obtained by Monte Carlo simulations of the model Eq.~(\ref{HeisenbergModel}) at $T/\Jin=0.01$, for $\tau =1.1$, $1.3$  and $1.5$.  
}
\end{figure}

\begin{figure}[h!]
\includegraphics[angle=0.0,  %trim = 0mm 0mm 10mm 0mm, clip,
 width=7cm ] {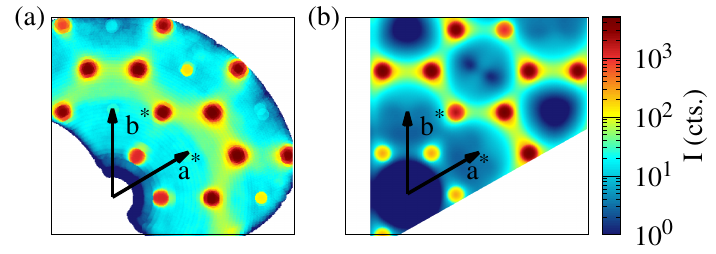}
\caption{\label{Fig1-magord} 
Comparison between the observed magnetic scattering  at 4\,K (a) and the results of the Monte Carlo  simulation of the Heisenberg model Eq.~(\ref{HeisenbergModel}) for $\tau =1.1$ (b).
 }
\end{figure}

%%%%%%%%%%%%%%%%%%%%%%%%%%
% DM interactions
\ifprintSections
\subsection{\DM interactions}
\fi
An important structural feature is the absence of inversion symmetry at the bonds connecting neighboring  magnetic ions, which gives rise to \DM interactions. 
According to the Keffer rule, \cite{Keffer1962} the direction of the DM vector $\mathbf D_{12}$, describing the strength and form of the interaction, is given by  $(\mathbf R_O- \mathbf R_1) \times (\mathbf R_O- \mathbf R_2)$, where $\mathbf R_{1,2}$ and $\mathbf R_O$ are the coordinates of the two magnetic ions and the oxygen ion, respectively. 
The directions of the DM vectors for neighboring Kagome sites,  $\mathbf D_{\mathrm {in}}$, and for bonds connecting the triangular and Kagome sites,  $\mathbf D_{\mathrm {out}}$, are shown in  Fig.~\ref{exchange}.

The DM interactions between the Kagome sites are likely very similar for intra- and inter-column neighbors. 
All of them have a component in the $ab$-plane depending on bond direction and a (larger) component along the $c$ axis.
The component of $\mathbf D_{\mathrm {in}}$ parallel to the $c$-axis gives rise to an easy-plane magnetic anisotropy of the Kagome spins.
However, along each row of bonds in the Kagome sublattice, the DM interactions have a common component parallel to $ab$-plane pointing perpendicular to the bond direction, which favors incommensurate cycloidal spirals with wave vectors in this plane. 
Due to the three-fold symmetry, the spiral wave vector can have three different directions. 
This can explain the short-range cycloidal spin correlations obeying the threefold symmetry, which were found using the polarization analysis of diffuse scattering. 
Particularly, in case of a long-range ordered spiral structure, the preservation of threefold symmetry is expected to result in a triple-q state.      

The DM interactions for out-of-plane bonds along the bipyramidal columns all lie in the $ab$-plane,  however they alternate in rotational sense for upward and downward paths from the trigonal to Kagome sites. 
Alternating  the sign of the DM interactions along the $c$-direction may create a spin canting between the Kagome and trigonal sites, which could possibly lead to weak ferromagnetism (see also Cheong {\em et al.}\cite{Cheong2007}).  
The effect could be subtle and depends on the actual magnetic ordered structure  and on the site anisotropies. 
The previously reported split positions of the oxygen ions\cite{Reim2014} result in small stochastic variations of the DM-vectors located along the corresponding bonds, which have been neglected in the following discussion.
%%%%%%%%%%%%%%%%%%%%%%%%%

%%%%%%%%%%%%%%%%%%%%%%%%%%
% Symmetry analysis of incommensurate ordering
\ifprintSections
\subsection{Symmetry analysis of incommensurate ordering}
\fi

To understand the nature of modulated states in the hexagonal swedenborgite, we first discuss symmetry properties of commensurate states with the wave vector ${\bf q}_c = (\nicefrac{1}{3},\,\nicefrac{1}{3},\,0)$. 
Symmetry analysis of magnetic ordering simplifies, if we assume that spins in blocks containing 8 transition metal sites (two spin triangles in Kagome layers and two connecting them spins in the triangular layers) are all collinear. 
A commensurate magnetic ordering with  the wave vector ${\bf q}_c$ contains three such blocks with the total spins ${\bf S}_{B1}, {\bf S}_{B2}$ and ${\bf S}_{B3}$ (see Fig.~\ref{fig:sitesandaxes}). 
Due to antiferromagnetic interactions between the blocks, these three spins add to zero: ${\bf S}_{B1} +{\bf S}_{B2} + {\bf S}_{B3} = 0$. 
Such states are described by $3\times3 - 3 = 6$ parameters, $X_1,Y_1,Z_1,X_2,Y_2,Z_2$:
\begin{align}
{\bf S}_{B1} &= (X_1,Y_1,Z_1), \nonumber\\
{\bf S}_{B2} &= (-\frac{1}{2} X_1 +\frac{\sqrt{3}}{2} X_2, -\frac{1}{2} Y_1 + \frac{\sqrt{3}}{2} Y_2, -\frac{1}{2} Z_1 + \frac{\sqrt{3}}{2} Z_2),\\
{\bf S}_{B3} &= (-\frac{1}{2} X_1 - \frac{\sqrt{3}}{2} X_2, -\frac{1}{2} Y_1 - \frac{\sqrt{3}}{2} Y_2, -\frac{1}{2} Z_1 - \frac{\sqrt{3}}{2} Z_2). \nonumber
\end{align}

\begin{figure}[htbp]
\includegraphics[width= 0.8\columnwidth]{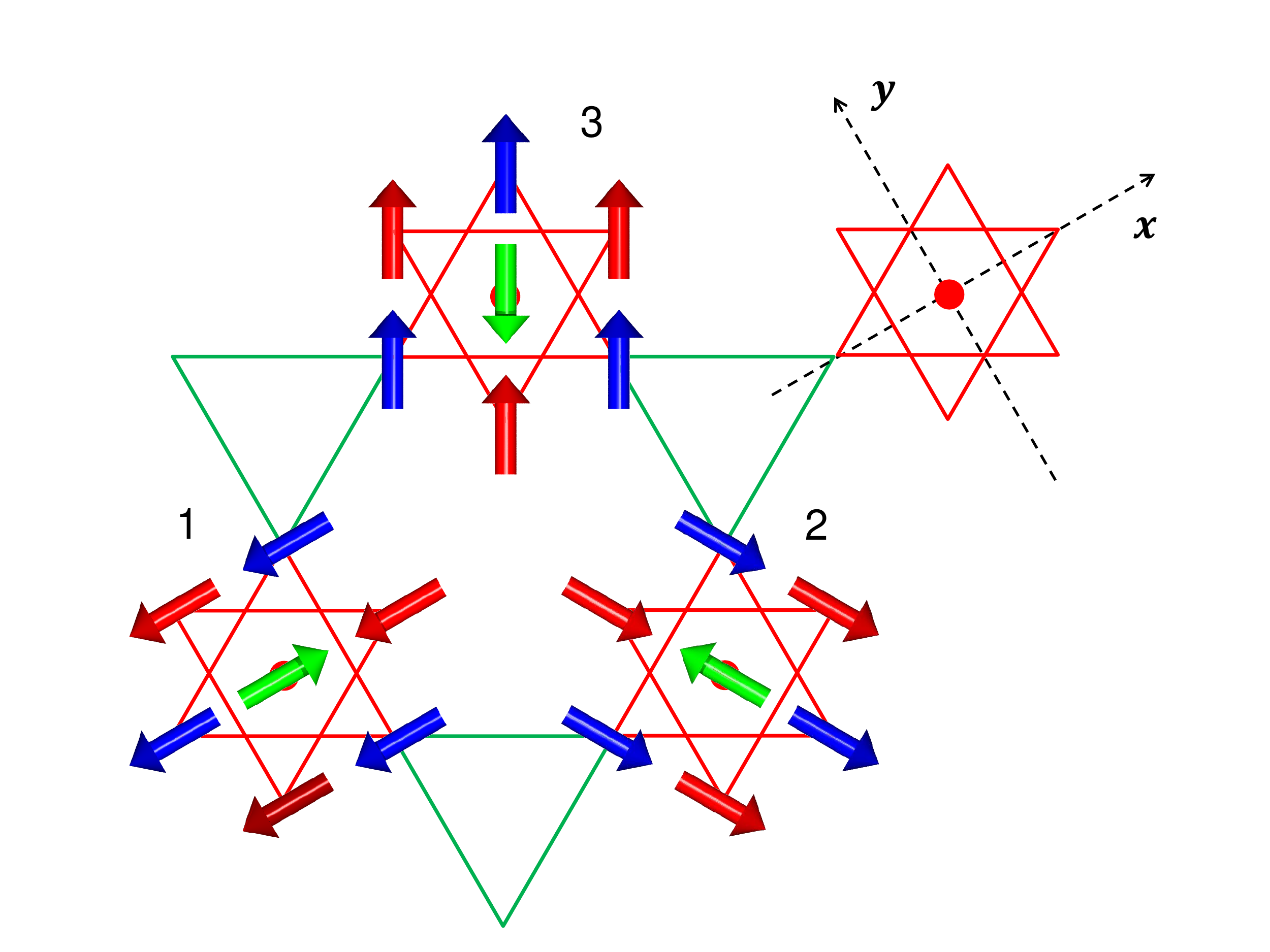}
\caption{\label{fig:sitesandaxes} 
Three magnetic building blocks with the total spins ${\bf S}_{B1}$, ${\bf S}_{B2}$ and ${\bf S}_{B3}$, forming the 120$^{\circ}$-state.
$x$ and $y$ denote the Cartesian axes in the $ab$-plane.)
}
\end{figure}

Transformation properties of these six parameters under the generators of the $P6_3mc$ space group are summarized in Table~\ref{tab:transformationsorders}, using which one can find two Lifshitz invariants resulting from the lack of inversion symmetry in the crystal lattice,
\begin{equation}
Z_1 \overset{\leftrightarrow}{\partial_x} X_1 + Z_2 \overset{\leftrightarrow}{\partial_x} X_2 + 
Z_1 \overset{\leftrightarrow}{\partial_y} Y_1 + Z_2 \overset{\leftrightarrow}{\partial_y} Y_2
\label{eq:LI1}
\end{equation}
and
\begin{equation}
X_1 \overset{\leftrightarrow}{\partial_x} Y_2 - X_2 \overset{\leftrightarrow}{\partial_x} Y_1 + 
X_1 \overset{\leftrightarrow}{\partial_y} X_2 - Y_1 \overset{\leftrightarrow}{\partial_y} Y_2,
\label{eq:LI2}
\end{equation}
where $A  \overset{\leftrightarrow}{\partial_i} B = A \partial_i B - B \partial_i A$, $i = x,y$.

The first Lifshitz invariant favors a cycloidal spiral in a vertical plane with the spin rotation axis, $\bf n$, normal to ${\bf S}_{B1}, {\bf S}_{B2}$ and ${\bf S}_{B3}$. 
The number of parameters describing a commensurate $120^\circ$-ordering of block is reduced to 4 by requiring ${\bf S}_{B1}^2 = {\bf S}_{B2}^2 = {\bf S}_{B3}^2$, which is equivalent to
\begin{equation}
\left\{
\begin{array}{l}
X_1^2 + Y_1^2 + Z_1^2 = X_2^2 + Y_2^2 + Z_2^2 \\ \\ 
X_1 X_2 + Y_1 Y_2 + Z_1 Z_2 = 0,
\end{array}
\right.
\end{equation}
i.e. the vectors ${\bf V}_1 = (X_1,Y_1,Z_1)$ and  ${\bf V}_2 = (X_2,Y_2,Z_2)$ have the same length and are orthogonal to each other. 
The Lifshitz invariant Eq.~(\ref{eq:LI1}) can then be written in the form
\begin{equation}
\sum_{i=x,y} (V_{1z} \overset{\leftrightarrow}{\partial_i} V_{1i} + V_{2z} \overset{\leftrightarrow}{\partial_i} V_{2i}).
\label{eq:LI1a}
\end{equation}
This invariant is similar to the one favoring the vertical cycloidal spiral in the polar hexagonal compound GaV$_4$Se$_8$\cite{Kezsmarki2015}, except that in GaV$_4$Se$_8$ the incommensurate magnetic state is locally close to the uniform ferromagnetic state, whereas in the swedenborgite it is close to the 120$^\circ$ antiferromagnetic ordering of the block spins. 
The cycloidal spiral favored by the Lifshitz invariant Eq.~(\ref{eq:LI1a}) preserves the norms of ${\bf V}_1$ and ${\bf V}_2$ as well as their orthogonality, as it favors a simultaneous rotation of  these two vectors around the axis ${\bf n} \parallel [{\bf V}_1 \times {\bf V}_2]$. 
The direction of the spiral wave vector, $\delta {\bf Q}_s$, (counted from the K-point in the reciprocal state) is left undetermined by this Lifshitz invariant of first order in gradient. 
The expansion to third order in $\delta {\bf Q}_s$  gives rise to three energy minima in the reciprocal space, 
$ \delta {\bf Q}_{s,1}$, $ \delta {\bf Q}_{s,2}$, and $ \delta {\bf Q}_{s,3}$, such that  
$ \delta {\bf Q}_{s,1} + \delta {\bf Q}_{s,2} + \delta {\bf Q}_{s,3} = 0 $,  
in the vicinity of the K-point. 
We also note that the change of sign of the vector chirality of the 120$^\circ$-state, obtained by $X_2 \rightarrow -X_2$, $Y_2 \rightarrow -Y_2$, $Z_2 \rightarrow - Z_2$, does not affect the direction of the spiral wave vector, as it leaves Eq.~(\ref{eq:LI1a}) unchanged.

The second Lifshitz invariant Eq.~(\ref{eq:LI2}) does not induce a modulated magnetic state with the local 120$^{\circ}$-structure: for two orthogonal in-plane vectors, ${\bf V}_1 = (X_1,Y_1,0)$ and ${\bf V}_2 = (-Y_1,X_1,0)$, this Lifshitz invariant is identically 0. 
More generally, if the condition ${\bf S}_{B1} + {\bf S}_{B2} + {\bf S}_{B3} = 0$ is relaxed, the analogue of Eq.~(\ref{eq:LI2}) is
\begin{align}
S_{B1}^x \overset{\leftrightarrow}{\partial_x} S_{B2}^y +& S_{B1}^y \overset{\leftrightarrow}{\partial_x} S_{B2}^x 
+ S_{B1}^x \overset{\leftrightarrow}{\partial_y} S_{B2}^x -  S_{B1}^y \overset{\leftrightarrow}{\partial_y} S_{B2}^y \nonumber\\
+ S_{B2}^x \overset{\leftrightarrow}{\partial_x} S_{B3}^y +& S_{B2}^y \overset{\leftrightarrow}{\partial_x} S_{B3}^x 
+ S_{B2}^x \overset{\leftrightarrow}{\partial_y} S_{B3}^x -  S_{B2}^y \overset{\leftrightarrow}{\partial_y} S_{B3}^y \\
+ S_{B3}^x \overset{\leftrightarrow}{\partial_x} S_{B1}^y +& S_{B3}^y \overset{\leftrightarrow}{\partial_x} S_{B1}^x 
+ S_{B3}^x \overset{\leftrightarrow}{\partial_y} S_{B1}^x -  S_{B3}^y \overset{\leftrightarrow}{\partial_y} S_{B1}^y. \nonumber
\end{align}
Also this Lifshitz invariant vanishes for a spiral state, in which ${\bf S}_{B1}$, ${\bf S}_{B2}$ and ${\bf S}_{B3}$ rotate in the $ab$-plane with the same wave vector. 
Thus we showed that the lack of inversion symmetry in the crystal lattice of the hexagonal swedenborgite induces a cycloidal spiral with spins rotating in a vertical plane and the wave vector parallel to the $ab$ plane, and does not favor spirals with spins rotating in the $ab$ plane.

Finally, we discuss a `non-linear weak ferromagnetism' of the 120$^{\circ}$-structures. 
Invariance under translations in the $ab$ plane and time reversal imply that the lowest power of the antiferromagnetic order parameter that can be coupled to a uniform magnetic field is 3. 
The lowest-order coupling allowed by symmetry is, 
\begin{equation}
H_c (Z_1^3 - 3 Z_1 Z_2^2).
\end{equation}
Only the antiferromagnetic states with spins out of the $ab$-plane can have a net ferromagnetic moment along the $c$ axis (another mechanism was discussed in Ref.~\onlinecite{Khalyavin2012}).
Net magnetization can be induced by multiply-periodic states, i.e. by the three coexisting minimal-energy cycloidal spirals with the wave vectors 
$\delta {\bf Q}_{s,1}$,   $\delta {\bf Q}_{s,2}$ and  $\delta {\bf Q}_{s,3}$ counted from the $K$-point ($\delta {\bf Q}_{s,1} + \delta {\bf Q}_{s,2} + \delta {\bf Q}_{s,3} = 0$).

\onecolumngrid

\begin{table}[!ht]
\centering
\begin{tabular}{|c|c|c|c|c|}
\hline
&$t_{a}$&$3_c$&$2_1$&$m_{a+b}$\\
\hline
$\left(\begin{array}{c}X_1\\X_2\\Y_1\\Y_2\end{array}\right)$&$\left(\begin{array}{cccc}-1/2 &-\sqrt{3}/2&0&0\\\sqrt{3}/{2}&-1/2&0&0\\0&0&-1/2&-\sqrt{3}/2\\0&0&\sqrt{3}/2&-1/2\end{array}\right)$&$\left(\begin{array}{cccc}-1/2 &0&-\sqrt{3}/2 &0\\0&-1/2&0&-\sqrt{3}/{2}\\\sqrt{3}/2&0&-1/2&0\\0&\sqrt{3}/2&0&-1/2\end{array}\right)$&$\left(\begin{array}{cccc}-1 &0&0&0\\0&1&0&0\\0&0&-1&0\\0&0&0&1\end{array}\right)$&$\left(\begin{array}{cccc}-1 &0&0&0\\0&1&0&0\\0&0&1&0\\0&0&0&-1\end{array}\right)$\\ &&&&\\
$\left(\begin{array}{c}Z_1\\Z_2\end{array}\right) $&$ \left( \begin{array}{cc} -1/2&-\sqrt{3}/2\\\sqrt{3}/2&-1/2\end{array}\right)$&$\left(\begin{array}{cc}1&0\\0&1\end{array}\right)$&$\left(\begin{array}{cc}1&0\\0&-1\end{array}\right)$&$\left(\begin{array}{cc}-1&0\\0&1\end{array}\right)$\\ &&&&\\
$\left(\begin{array}{c}\partial_x\\\partial_y\end{array}\right) $&$ \left( \begin{array}{cc} 1&0\\0&1\end{array}\right)$&$\left(\begin{array}{cc}-1/2&-\sqrt{3}/2\\\sqrt{3}/2&-1/2\end{array}\right)$&$\left(\begin{array}{cc}-1&0\\0&-1\end{array}\right)$&$\left(\begin{array}{cc}1&0\\0&-1\end{array}\right)$\\
[0.4ex]
\hline
\end{tabular}
\caption{Transformations of the six order parameters and the in-plane components of gradient under the generators of the P$6_3mc$ space group: translation $t_a = (x+1,y,z)$, three-fold axis  $3_c = (-y,x-y,z)$, two-fold screw axis ${2}_1 = (-x,-y,z+1/2)$ and mirror plane $m_{a+b} = (-y,-x,z)$.}
\label{tab:transformationsorders}
\end{table}

\twocolumngrid

%%%%%%%%%%%%%%%%%%%%%%%%%%
% Conclusion
\ifprintSections  
\section{ Summary and Discussion }
\fi

% % % % % % % % % % % % % % % % % % % % % % % % 
To summarize,  \Ca shows a unique coexistence of the long-range {\ExE}antiferromagnetic order and partial disorder. 
The latter results from geometric frustration in the layered Kagome lattice, which leads to ground state degeneracy and diffuse scattering.
Surprisingly, the peak structure shows a great deal of coherence corresponding to an incipient incommensurate spin-spiral order. 
We argue that these correlations result from the lack of inversion symmetry in the swedenborgite crystal lattice and the concomitant Lifshitz invariant favoring spiral states.   

Our neutron diffraction experiments on a single crystal of \Ca  with the layered Kagome structure provide a detailed insight into the complex non-collinear spin structure of  this geometrically frustrated swedenborgite.
In the first approximation, magnetic properties of \Ca are described by the nearest-neighbor Heisenberg model Eq.~(\ref{HeisenbergModel}) with relatively strong antiferromagnetic superexchange interactions. 
Monte Carlo simulations of this model compare well with the observed distribution of the diffuse scattering intensity in the reciprocal space. 
The best fit is obtained for $\tau = \frac{\Jout}{\Jin} \approx 1.1$, corresponding to an extensive ground state degeneracy of the model, which is consistent with the observed reduction of ordered magnetic moments and with the constraints, Eqs.~(\ref{eq:Striangle}) and (\ref{eq:striag}), first discussed in Ref.~\onlinecite{Manuel2009}. 
We considered in detail spiral ground states and showed that, for $\tau < 1.5$, the spiral wave vector can vary in a circular-shaped region around the K point without changing the energy of the spiral state. 
This flat energy landscape in the reciprocal state gives rise to the observed intensity distribution of diffuse scattering.

The simple Heisenberg model cannot, however, describe the structure of the magnetic peaks and the complex temperature dependence of the magnetic order that sets in near $T_\textrm{N}=160$\,K.
Neutron diffraction, polarization analysis and magnetization measurements show that near the transition temperature ordered spins lie in the $ab$-plane. 
However, they cant out of this plane as temperature decreases.
The same conclusion was reached  on the basis of  the  magnetic structure determination (see Table~\ref{SASolution}) that does not make use of the polarization analysis.
The best refinement is found within the trigonal magnetic space groups, $P31m'$ and $P31m$, the former group being the most probable solution.
The reduced ordered moment from the refinement is in agreement with the partial order.

Polarization analysis of the magnetic scattering revealed the chiral nature of spin correlations, which are of the cycloidal type with both the spiral wave vector, ${\mathbf Q}_s$, and the chirality vector (the spin rotation axis), $\mathbf C$,  in the $ab$-plane.
The cycloidal spiral with spins rotating around the $c$-axis in the $ab$-plane may also be present, but it is not visible in the $(hk0)$-scattering plane
 as only the spin components perpendicular to the scattering vector can be measured.
We can exclude, however, any significant contribution from the helical spirals in the $ab$-plane.
Polarization analysis of the $(hhl)$-scattering plane also excludes within experimental resolution any long-period
spiral spin correlations (cycloidal or helical) out of the $ab$-plane.

The scattering near the $K$-point is consistent with our theoretical analysis of degenerate spiral states in the Heisenberg model.
In addition, the scattering from the $c$-components of spins indicates a slight preference in the direction of the wave vector of the incipient incommensurate spiral states, corresponding to  three energy minima in the reciprocal space near the $K$-points.
Using symmetry analysis we have found a Lifshitz invariant that favors an incommensurate cycloidal spiral state in which the 120$^\circ$-order slowly rotates around an axis in the $ab$-plane.
The tendency towards an incommensurate spiral modulation originates from the \DM interaction  between the spins of the transition metal ions in the tetrahedral oxygen coordination and the lack of inversion symmetry in the swedenborgite crystal lattice.
The same mechanism stabilizes spiral and skyrmion crystal states in MnSi, Cu$_2$OSeO$_3$ and other chiral ferromagnets \cite{Roessler2006,Butenko2010,Okubo2012}, the difference being that in the swedenborgite we find the rotation of the {\ExE} antiferromagnetic spin order.

The \DM interaction added to the frustrated Heisenberg model selects (1) the cycloidal type of spiral ordering and (2) three energy minima in the vicinity of the $K$-point, corresponding to three spiral states with the 120$^\circ$-angle between the wave vectors in the $ab$-plane.
Our symmetry analysis shows that the simultaneous presence of three spiral orders can induce a weak ferromagnetic moment found in our magnetization measurements.
%
%The three coexisting spirals would form an antiferromagnetic skyrmion crystal.
One possible multiply-periodic state with three coexisting spirals is an antiferromagnetic skyrmion crystal.
Thus swedenborgites with the crystal lattice that is both geometrically frustrated and chiral may turn out to be a new interesting family of skyrmion materials.

% % % % % % % % % % % % % % % % % % % % % % % % 

\begin{acknowledgments}
We gratefully acknowledge Jochen Stahn and Simon Ward  for their help with the set up of the instrument Morpheus,
and Artur Glavic for support by providing the software plot.py for the  evaluation of the data measured at DNS. 
Furthermore, we thank Juan Rodr{\'\i}guez-Carvajal for his advice upon the magnetic structure refinement.
MM would like to thank D. D. Khalyavin for useful discussions and JDR thanks D. Okuyama for providing scientific support. 
%The neutron diffraction experiments were performed at the FRM-2 at the Maier-Leibniz Zentrum, Munich, Germany, and at the
%Swiss spallation neutron source SINQ, at Paul Scherrer Institute, Villigen, Switzerland.
This work is based on experiments performed at the Swiss spallation neutron source SINQ, Paul Scherrer Institute, Villigen, Switzerland and 
at the DNS instrument operated by JCNS at the Heinz Maier-Leibnitz Zentrum (MLZ), Garching, Germany.
This work was supported by the German Science Foundation (DFG) through SFB 1143.
Work at IMRAM was partly supported by the Research Program ``Dynamic Alliance for Open Innovation Bridging Human, Environment and Materials''.
JDR is an International Research Fellow  of the  Japan Society for the Promotion of Science.
\end{acknowledgments}

% \section*{References}
% K. Momma and F. Izumi, {\em VESTA 3 for three-dimensional visualization of crystal, volumetric and morphology data}, J. Appl. Crystallogr., 44, 1272-1276 (2011).
%\bibliography{complete.bib}

\begin{thebibliography}{36}%
\makeatletter
\providecommand \@ifxundefined [1]{%
 \@ifx{#1\undefined}
}%
\providecommand \@ifnum [1]{%
 \ifnum #1\expandafter \@firstoftwo
 \else \expandafter \@secondoftwo
 \fi
}%
\providecommand \@ifx [1]{%
 \ifx #1\expandafter \@firstoftwo
 \else \expandafter \@secondoftwo
 \fi
}%
\providecommand \natexlab [1]{#1}%
\providecommand \enquote  [1]{``#1''}%
\providecommand \bibnamefont  [1]{#1}%
\providecommand \bibfnamefont [1]{#1}%
\providecommand \citenamefont [1]{#1}%
\providecommand \href@noop [0]{\@secondoftwo}%
\providecommand \href [0]{\begingroup \@sanitize@url \@href}%
\providecommand \@href[1]{\@@startlink{#1}\@@href}%
\providecommand \@@href[1]{\endgroup#1\@@endlink}%
\providecommand \@sanitize@url [0]{\catcode `\\12\catcode `\$12\catcode
  `\&12\catcode `\#12\catcode `\^12\catcode `\_12\catcode `\%12\relax}%
\providecommand \@@startlink[1]{}%
\providecommand \@@endlink[0]{}%
\providecommand \url  [0]{\begingroup\@sanitize@url \@url }%
\providecommand \@url [1]{\endgroup\@href {#1}{\urlprefix }}%
\providecommand \urlprefix  [0]{URL }%
\providecommand \Eprint [0]{\href }%
\providecommand \doibase [0]{http://dx.doi.org/}%
\providecommand \selectlanguage [0]{\@gobble}%
\providecommand \bibinfo  [0]{\@secondoftwo}%
\providecommand \bibfield  [0]{\@secondoftwo}%
\providecommand \translation [1]{[#1]}%
\providecommand \BibitemOpen [0]{}%
\providecommand \bibitemStop [0]{}%
\providecommand \bibitemNoStop [0]{.\EOS\space}%
\providecommand \EOS [0]{\spacefactor3000\relax}%
\providecommand \BibitemShut  [1]{\csname bibitem#1\endcsname}%
\let\auto@bib@innerbib\@empty
%</preamble>
\bibitem [{\citenamefont {Ramirez}(1994)}]{Ramirez1994}%
  \BibitemOpen
  \bibfield  {author} {\bibinfo {author} {\bibfnamefont {A.~P.}\ \bibnamefont
  {Ramirez}},\ }\href@noop {} {\bibfield  {journal} {\bibinfo  {journal}
  {Annual Review of Materials Science}\ }\textbf {\bibinfo {volume} {24}},\
  \bibinfo {pages} {453} (\bibinfo {year} {1994})}\BibitemShut {NoStop}%
\bibitem [{\citenamefont {Moessner}\ and\ \citenamefont
  {Ramirez}(2006)}]{Moessner2006}%
  \BibitemOpen
  \bibfield  {author} {\bibinfo {author} {\bibfnamefont {R.}~\bibnamefont
  {Moessner}}\ and\ \bibinfo {author} {\bibfnamefont {A.~R.}\ \bibnamefont
  {Ramirez}},\ }\href@noop {} {\bibfield  {journal} {\bibinfo  {journal}
  {Physics Today}\ }\textbf {\bibinfo {volume} {59}},\ \bibinfo {pages} {24}
  (\bibinfo {year} {2006})}\BibitemShut {NoStop}%
\bibitem [{\citenamefont {Dzyaloshinskii}(1964)}]{Dzyaloshinskii1964}%
  \BibitemOpen
  \bibfield  {author} {\bibinfo {author} {\bibfnamefont {I.~E.}\ \bibnamefont
  {Dzyaloshinskii}},\ }\href@noop {} {\bibfield  {journal} {\bibinfo  {journal}
  {Soviet Physics Jetp-ussr}\ }\textbf {\bibinfo {volume} {19}},\ \bibinfo
  {pages} {960} (\bibinfo {year} {1964})}\BibitemShut {NoStop}%
\bibitem [{\citenamefont {Moriya}(1960)}]{Moriya1960}%
  \BibitemOpen
  \bibfield  {author} {\bibinfo {author} {\bibfnamefont {T.}~\bibnamefont
  {Moriya}},\ }\href@noop {} {\bibfield  {journal} {\bibinfo  {journal}
  {Physical Review}\ }\textbf {\bibinfo {volume} {120}},\ \bibinfo {pages} {91}
  (\bibinfo {year} {1960})}\BibitemShut {NoStop}%
\bibitem [{\citenamefont {Pauling}\ \emph {et~al.}(1935)\citenamefont
  {Pauling}, \citenamefont {King},\ and\ \citenamefont
  {Winchell}}]{Pauling1935}%
  \BibitemOpen
  \bibfield  {author} {\bibinfo {author} {\bibfnamefont {L.}~\bibnamefont
  {Pauling}}, \bibinfo {author} {\bibfnamefont {H.~P.}\ \bibnamefont {King}}, \
  and\ \bibinfo {author} {\bibfnamefont {A.~N.}\ \bibnamefont {Winchell}},\
  }\href@noop {} {\bibfield  {journal} {\bibinfo  {journal} {American
  Mineralogist}\ }\textbf {\bibinfo {volume} {20}},\ \bibinfo {pages} {402}
  (\bibinfo {year} {1935})}\BibitemShut {NoStop}%
\bibitem [{\citenamefont {Valldor}\ and\ \citenamefont
  {Andersson}(2002)}]{Valldor2002}%
  \BibitemOpen
  \bibfield  {author} {\bibinfo {author} {\bibfnamefont {M.}~\bibnamefont
  {Valldor}}\ and\ \bibinfo {author} {\bibfnamefont {M.}~\bibnamefont
  {Andersson}},\ }\href@noop {} {\bibfield  {journal} {\bibinfo  {journal}
  {Solid State Sciences}\ }\textbf {\bibinfo {volume} {4}},\ \bibinfo {pages}
  {923} (\bibinfo {year} {2002})}\BibitemShut {NoStop}%
\bibitem [{\citenamefont {Valldor}(2004)}]{Valldor2004a}%
  \BibitemOpen
  \bibfield  {author} {\bibinfo {author} {\bibfnamefont {M.}~\bibnamefont
  {Valldor}},\ }\href {\doibase 10.1016/j.solidstatesciences.2004.01.004}
  {\bibfield  {journal} {\bibinfo  {journal} {Solid State Sciences}\ }\textbf
  {\bibinfo {volume} {6}},\ \bibinfo {pages} {251} (\bibinfo {year}
  {2004})}\BibitemShut {NoStop}%
\bibitem [{\citenamefont {Valldor}\ \emph {et~al.}(2009)\citenamefont
  {Valldor}, \citenamefont {Sanders},\ and\ \citenamefont
  {Schweika}}]{Valldor2009}%
  \BibitemOpen
  \bibfield  {author} {\bibinfo {author} {\bibfnamefont {M.}~\bibnamefont
  {Valldor}}, \bibinfo {author} {\bibfnamefont {Y.}~\bibnamefont {Sanders}}, \
  and\ \bibinfo {author} {\bibfnamefont {W.}~\bibnamefont {Schweika}},\ }\href
  {\doibase 10.1088/1742-6596/145/1/012076} {\bibfield  {journal} {\bibinfo
  {journal} {Journal of Physics: Conference Series}\ }\textbf {\bibinfo
  {volume} {145}},\ \bibinfo {pages} {012076} (\bibinfo {year}
  {2009})}\BibitemShut {NoStop}%
\bibitem [{\citenamefont {Manuel}\ \emph {et~al.}(2009)\citenamefont {Manuel},
  \citenamefont {Chapon}, \citenamefont {Radaelli}, \citenamefont {Zheng},\
  and\ \citenamefont {Mitchell}}]{Manuel2009}%
  \BibitemOpen
  \bibfield  {author} {\bibinfo {author} {\bibfnamefont {P.}~\bibnamefont
  {Manuel}}, \bibinfo {author} {\bibfnamefont {L.~C.}\ \bibnamefont {Chapon}},
  \bibinfo {author} {\bibfnamefont {P.~G.}\ \bibnamefont {Radaelli}}, \bibinfo
  {author} {\bibfnamefont {H.}~\bibnamefont {Zheng}}, \ and\ \bibinfo {author}
  {\bibfnamefont {J.~F.}\ \bibnamefont {Mitchell}},\ }\href {\doibase
  10.1103/PhysRevLett.103.037202} {\bibfield  {journal} {\bibinfo  {journal}
  {Physical Review Letters}\ }\textbf {\bibinfo {volume} {103}},\ \bibinfo
  {pages} {037202} (\bibinfo {year} {2009})}\BibitemShut {NoStop}%
\bibitem [{\citenamefont {Khalyavin}\ \emph {et~al.}(2010)\citenamefont
  {Khalyavin}, \citenamefont {Manuel}, \citenamefont {Mitchell},\ and\
  \citenamefont {Chapon}}]{Khalyavin2010}%
  \BibitemOpen
  \bibfield  {author} {\bibinfo {author} {\bibfnamefont {D.~D.}\ \bibnamefont
  {Khalyavin}}, \bibinfo {author} {\bibfnamefont {P.}~\bibnamefont {Manuel}},
  \bibinfo {author} {\bibfnamefont {J.~F.}\ \bibnamefont {Mitchell}}, \ and\
  \bibinfo {author} {\bibfnamefont {L.~C.}\ \bibnamefont {Chapon}},\ }\href
  {\doibase 10.1103/PhysRevB.82.094401} {\bibfield  {journal} {\bibinfo
  {journal} {Physical Review B}\ }\textbf {\bibinfo {volume} {82}},\ \bibinfo
  {pages} {094401} (\bibinfo {year} {2010})}\BibitemShut {NoStop}%
\bibitem [{\citenamefont {Valldor}\ \emph {et~al.}(2011)\citenamefont
  {Valldor}, \citenamefont {Hermann}, \citenamefont {Wuttke}, \citenamefont
  {Zamponi},\ and\ \citenamefont {Schweika}}]{Valldor2011}%
  \BibitemOpen
  \bibfield  {author} {\bibinfo {author} {\bibfnamefont {M.}~\bibnamefont
  {Valldor}}, \bibinfo {author} {\bibfnamefont {R.~P.}\ \bibnamefont
  {Hermann}}, \bibinfo {author} {\bibfnamefont {J.}~\bibnamefont {Wuttke}},
  \bibinfo {author} {\bibfnamefont {M.}~\bibnamefont {Zamponi}}, \ and\
  \bibinfo {author} {\bibfnamefont {W.}~\bibnamefont {Schweika}},\ }\href
  {\doibase 10.1103/PhysRevB.84.224426} {\bibfield  {journal} {\bibinfo
  {journal} {Physical Review B}\ }\textbf {\bibinfo {volume} {84}},\ \bibinfo
  {pages} {224426} (\bibinfo {year} {2011})}\BibitemShut {NoStop}%
\bibitem [{\citenamefont {Schweika}\ \emph {et~al.}(2007)\citenamefont
  {Schweika}, \citenamefont {Valldor},\ and\ \citenamefont
  {Lemmens}}]{Schweika2007}%
  \BibitemOpen
  \bibfield  {author} {\bibinfo {author} {\bibfnamefont {W.}~\bibnamefont
  {Schweika}}, \bibinfo {author} {\bibfnamefont {M.}~\bibnamefont {Valldor}}, \
  and\ \bibinfo {author} {\bibfnamefont {P.}~\bibnamefont {Lemmens}},\ }\href
  {\doibase 10.1103/PhysRevLett.98.067201} {\bibfield  {journal} {\bibinfo
  {journal} {Physical Review Letters}\ }\textbf {\bibinfo {volume} {98}},\
  \bibinfo {pages} {067201} (\bibinfo {year} {2007})}\BibitemShut {NoStop}%
\bibitem [{\citenamefont {Stewart}\ \emph {et~al.}(2011)\citenamefont
  {Stewart}, \citenamefont {Ehlers}, \citenamefont {Mutka}, \citenamefont
  {Fouquet}, \citenamefont {Payen},\ and\ \citenamefont {Lortz}}]{Stewart2011}%
  \BibitemOpen
  \bibfield  {author} {\bibinfo {author} {\bibfnamefont {J.~R.}\ \bibnamefont
  {Stewart}}, \bibinfo {author} {\bibfnamefont {G.}~\bibnamefont {Ehlers}},
  \bibinfo {author} {\bibfnamefont {H.}~\bibnamefont {Mutka}}, \bibinfo
  {author} {\bibfnamefont {P.}~\bibnamefont {Fouquet}}, \bibinfo {author}
  {\bibfnamefont {C.}~\bibnamefont {Payen}}, \ and\ \bibinfo {author}
  {\bibfnamefont {R.}~\bibnamefont {Lortz}},\ }\href {\doibase
  10.1103/PhysRevB.83.024405} {\bibfield  {journal} {\bibinfo  {journal}
  {Physical Review B}\ }\textbf {\bibinfo {volume} {83}},\ \bibinfo {pages}
  {024405} (\bibinfo {year} {2011})}\BibitemShut {NoStop}%
\bibitem [{\citenamefont {Reim}\ \emph {et~al.}(2014)\citenamefont {Reim},
  \citenamefont {Rosen}, \citenamefont {Schweika}, \citenamefont {Meven},
  \citenamefont {Leo}, \citenamefont {Meier}, \citenamefont {Fiebig},
  \citenamefont {Schmidt}, \citenamefont {Kuo}, \citenamefont {Pi},
  \citenamefont {Hu},\ and\ \citenamefont {Valldor}}]{Reim2014}%
  \BibitemOpen
  \bibfield  {author} {\bibinfo {author} {\bibfnamefont {J.~D.}\ \bibnamefont
  {Reim}}, \bibinfo {author} {\bibfnamefont {E.}~\bibnamefont {Rosen}},
  \bibinfo {author} {\bibfnamefont {W.}~\bibnamefont {Schweika}}, \bibinfo
  {author} {\bibfnamefont {M.}~\bibnamefont {Meven}}, \bibinfo {author}
  {\bibfnamefont {N.~R.}\ \bibnamefont {Leo}}, \bibinfo {author} {\bibfnamefont
  {D.}~\bibnamefont {Meier}}, \bibinfo {author} {\bibfnamefont
  {M.}~\bibnamefont {Fiebig}}, \bibinfo {author} {\bibfnamefont
  {M.}~\bibnamefont {Schmidt}}, \bibinfo {author} {\bibfnamefont {C.-Y.}\
  \bibnamefont {Kuo}}, \bibinfo {author} {\bibfnamefont {T.-W.}\ \bibnamefont
  {Pi}}, \bibinfo {author} {\bibfnamefont {Z.}~\bibnamefont {Hu}}, \ and\
  \bibinfo {author} {\bibfnamefont {M.}~\bibnamefont {Valldor}},\ }\href
  {\doibase 10.1107/S1600576714023528} {\bibfield  {journal} {\bibinfo
  {journal} {Journal of Applied Crystallography}\ }\textbf {\bibinfo {volume}
  {47}},\ \bibinfo {pages} {2038} (\bibinfo {year} {2014})}\BibitemShut
  {NoStop}%
\bibitem [{\citenamefont {Buhrandt}\ and\ \citenamefont
  {Fritz}(2014)}]{Buhrandt2014}%
  \BibitemOpen
  \bibfield  {author} {\bibinfo {author} {\bibfnamefont {S.}~\bibnamefont
  {Buhrandt}}\ and\ \bibinfo {author} {\bibfnamefont {L.}~\bibnamefont
  {Fritz}},\ }\href {\doibase 10.1103/PhysRevB.90.020403} {\bibfield  {journal}
  {\bibinfo  {journal} {Physical Review B}\ }\textbf {\bibinfo {volume} {90}},\
  \bibinfo {pages} {020403} (\bibinfo {year} {2014})}\BibitemShut {NoStop}%
\bibitem [{\citenamefont {Reimers}\ and\ \citenamefont
  {Berlinsky}(1993)}]{Reimers1993}%
  \BibitemOpen
  \bibfield  {author} {\bibinfo {author} {\bibfnamefont {J.~N.}\ \bibnamefont
  {Reimers}}\ and\ \bibinfo {author} {\bibfnamefont {A.~J.}\ \bibnamefont
  {Berlinsky}},\ }\href@noop {} {\bibfield  {journal} {\bibinfo  {journal}
  {Physical Review B}\ }\textbf {\bibinfo {volume} {48}},\ \bibinfo {pages}
  {9539} (\bibinfo {year} {1993})}\BibitemShut {NoStop}%
\bibitem [{\citenamefont {Zhitomirsky}(2008)}]{Zhitomirsky2008}%
  \BibitemOpen
  \bibfield  {author} {\bibinfo {author} {\bibfnamefont {M.~E.}\ \bibnamefont
  {Zhitomirsky}},\ }\href {\doibase 10.1103/PhysRevB.78.094423} {\bibfield
  {journal} {\bibinfo  {journal} {Physical Review B}\ }\textbf {\bibinfo
  {volume} {78}},\ \bibinfo {pages} {094423} (\bibinfo {year}
  {2008})}\BibitemShut {NoStop}%
\bibitem [{\citenamefont {Schweika}\ and\ \citenamefont
  {B\"oni}(2001)}]{Schweika2001}%
  \BibitemOpen
  \bibfield  {author} {\bibinfo {author} {\bibfnamefont {W.}~\bibnamefont
  {Schweika}}\ and\ \bibinfo {author} {\bibfnamefont {P.}~\bibnamefont
  {B\"oni}},\ }\href {\doibase 10.1016/S0921-4526(00)00858-9} {\bibfield
  {journal} {\bibinfo  {journal} {Physica B-Condensed Matter}\ }\textbf
  {\bibinfo {volume} {297}},\ \bibinfo {pages} {155} (\bibinfo {year}
  {2001})}\BibitemShut {NoStop}%
\bibitem [{\citenamefont {Sch\"arpf}\ and\ \citenamefont
  {Capellmann}(1993)}]{Scharpf1993}%
  \BibitemOpen
  \bibfield  {author} {\bibinfo {author} {\bibfnamefont {O.}~\bibnamefont
  {Sch\"arpf}}\ and\ \bibinfo {author} {\bibfnamefont {H.}~\bibnamefont
  {Capellmann}},\ }\href {\doibase 10.1002/pssa.2211350204} {\bibfield
  {journal} {\bibinfo  {journal} {Physica Status Solidi A-applied Research}\
  }\textbf {\bibinfo {volume} {135}},\ \bibinfo {pages} {359} (\bibinfo {year}
  {1993})}\BibitemShut {NoStop}%
\bibitem [{\citenamefont {Schweika}(2010)}]{Schweika2010}%
  \BibitemOpen
  \bibfield  {author} {\bibinfo {author} {\bibfnamefont {W.}~\bibnamefont
  {Schweika}},\ }\href {\doibase 10.1088/1742-6596/211/1/012026} {\bibfield
  {journal} {\bibinfo  {journal} {Journal of Physics Conference Series}\
  }\textbf {\bibinfo {volume} {211}},\ \bibinfo {pages} {012026} (\bibinfo
  {year} {2010})}\BibitemShut {NoStop}%
\bibitem [{\citenamefont {Perez-Mato}\ \emph {et~al.}(2015)\citenamefont
  {Perez-Mato}, \citenamefont {Gallego}, \citenamefont {Tasci}, \citenamefont
  {Elcoro}, \citenamefont {de~la Flor},\ and\ \citenamefont
  {Aroyo}}]{Perez-Mato2015}%
  \BibitemOpen
  \bibfield  {author} {\bibinfo {author} {\bibfnamefont {J.~M.}\ \bibnamefont
  {Perez-Mato}}, \bibinfo {author} {\bibfnamefont {S.~V.}\ \bibnamefont
  {Gallego}}, \bibinfo {author} {\bibfnamefont {E.~S.}\ \bibnamefont {Tasci}},
  \bibinfo {author} {\bibfnamefont {L.}~\bibnamefont {Elcoro}}, \bibinfo
  {author} {\bibfnamefont {G.}~\bibnamefont {de~la Flor}}, \ and\ \bibinfo
  {author} {\bibfnamefont {M.~I.}\ \bibnamefont {Aroyo}},\ }\href {\doibase
  10.1146/annurev-matsci-070214-021008} {\bibfield  {journal} {\bibinfo
  {journal} {Annual Review of Materials Research, Vol 45}\ }\textbf {\bibinfo
  {volume} {45}},\ \bibinfo {pages} {217} (\bibinfo {year} {2015})}\BibitemShut
  {NoStop}%
\bibitem [{\citenamefont
  {Rodr\'{i}guez-Carvajal}(1993)}]{RodriguezCaravajal1993}%
  \BibitemOpen
  \bibfield  {author} {\bibinfo {author} {\bibfnamefont {J.}~\bibnamefont
  {Rodr\'{i}guez-Carvajal}},\ }\href {\doibase 10.1016/0921-4526(93)90108-I}
  {\bibfield  {journal} {\bibinfo  {journal} {Physica B}\ }\textbf {\bibinfo
  {volume} {192}},\ \bibinfo {pages} {55} (\bibinfo {year} {1993})}\BibitemShut
  {NoStop}%
\bibitem [{\citenamefont {Villain}\ \emph {et~al.}(1980)\citenamefont
  {Villain}, \citenamefont {Bidaux}, \citenamefont {Carton},\ and\
  \citenamefont {Conte}}]{Villain1980}%
  \BibitemOpen
  \bibfield  {author} {\bibinfo {author} {\bibfnamefont {J.}~\bibnamefont
  {Villain}}, \bibinfo {author} {\bibfnamefont {R.}~\bibnamefont {Bidaux}},
  \bibinfo {author} {\bibfnamefont {J.~P.}\ \bibnamefont {Carton}}, \ and\
  \bibinfo {author} {\bibfnamefont {R.}~\bibnamefont {Conte}},\ }\href@noop {}
  {\bibfield  {journal} {\bibinfo  {journal} {Journal De Physique}\ }\textbf
  {\bibinfo {volume} {41}},\ \bibinfo {pages} {1263} (\bibinfo {year}
  {1980})}\BibitemShut {NoStop}%
\bibitem [{\citenamefont {Litvin}(2008)}]{Litvin2008}%
  \BibitemOpen
  \bibfield  {author} {\bibinfo {author} {\bibfnamefont {D.~B.}\ \bibnamefont
  {Litvin}},\ }\href {\doibase 10.1107/S010876730800768X} {\bibfield  {journal}
  {\bibinfo  {journal} {Acta Crystallographica Section A}\ }\textbf {\bibinfo
  {volume} {64}},\ \bibinfo {pages} {419} (\bibinfo {year} {2008})}\BibitemShut
  {NoStop}%
\bibitem [{Rei()}]{Reim2018}%
  \BibitemOpen
  \href@noop {} {}\bibinfo {note} {Work in preparation.}\BibitemShut {Stop}%
\bibitem [{\citenamefont {Chatterjee}\ and\ \citenamefont
  {Saha-Dasgupta}(2011)}]{Chatterjee2011}%
  \BibitemOpen
  \bibfield  {author} {\bibinfo {author} {\bibfnamefont {S.}~\bibnamefont
  {Chatterjee}}\ and\ \bibinfo {author} {\bibfnamefont {T.}~\bibnamefont
  {Saha-Dasgupta}},\ }\href {\doibase 10.1103/physrevb.84.085116} {\bibfield
  {journal} {\bibinfo  {journal} {Physical Review B}\ }\textbf {\bibinfo
  {volume} {84}},\ \bibinfo {pages} {085116} (\bibinfo {year}
  {2011})}\BibitemShut {NoStop}%
\bibitem [{\citenamefont {Goodenough}(1963)}]{Goodenough1963}%
  \BibitemOpen
  \bibfield  {author} {\bibinfo {author} {\bibfnamefont {J.~B.}\ \bibnamefont
  {Goodenough}},\ }\href@noop {} {\emph {\bibinfo {title} {Magnetism and the
  Chemical Bond}}}\ (\bibinfo  {publisher} {Interscience-Wiley},\ \bibinfo
  {year} {1963})\BibitemShut {NoStop}%
\bibitem [{\citenamefont {Anderson}(1963)}]{Anderson1963}%
  \BibitemOpen
  \bibfield  {author} {\bibinfo {author} {\bibfnamefont {P.~W.}\ \bibnamefont
  {Anderson}},\ }\href@noop {} {\bibfield  {journal} {\bibinfo  {journal}
  {Solid State Physics}\ }\textbf {\bibinfo {volume} {14}},\ \bibinfo {pages}
  {99} (\bibinfo {year} {1963})}\BibitemShut {NoStop}%
\bibitem [{\citenamefont {Khomskii}(2014)}]{Khomskii2014}%
  \BibitemOpen
  \bibfield  {author} {\bibinfo {author} {\bibfnamefont {D.~I.}\ \bibnamefont
  {Khomskii}},\ }\href@noop {} {\emph {\bibinfo {title} {Transition Metal
  Compounds}}}\ (\bibinfo  {publisher} {Cambridge University Press},\ \bibinfo
  {year} {2014})\BibitemShut {NoStop}%
\bibitem [{\citenamefont {Keffer}(1962)}]{Keffer1962}%
  \BibitemOpen
  \bibfield  {author} {\bibinfo {author} {\bibfnamefont {F.}~\bibnamefont
  {Keffer}},\ }\href@noop {} {\bibfield  {journal} {\bibinfo  {journal}
  {Physical Review}\ }\textbf {\bibinfo {volume} {126}},\ \bibinfo {pages}
  {896} (\bibinfo {year} {1962})}\BibitemShut {NoStop}%
\bibitem [{\citenamefont {Cheong}\ and\ \citenamefont
  {Mostovoy}(2007)}]{Cheong2007}%
  \BibitemOpen
  \bibfield  {author} {\bibinfo {author} {\bibfnamefont {S.-W.}\ \bibnamefont
  {Cheong}}\ and\ \bibinfo {author} {\bibfnamefont {M.}~\bibnamefont
  {Mostovoy}},\ }\href {\doibase 10.1038/nmat1804} {\bibfield  {journal}
  {\bibinfo  {journal} {Nature Materials}\ }\textbf {\bibinfo {volume} {6}},\
  \bibinfo {pages} {13} (\bibinfo {year} {2007})}\BibitemShut {NoStop}%
\bibitem [{\citenamefont {Kezsmarki}\ \emph {et~al.}(2015)\citenamefont
  {Kezsmarki}, \citenamefont {Bordacs}, \citenamefont {Milde}, \citenamefont
  {Neuber}, \citenamefont {Eng}, \citenamefont {White}, \citenamefont {Ronnow},
  \citenamefont {Dewhurst}, \citenamefont {Mochizuki}, \citenamefont {Yanai},
  \citenamefont {Nakamura}, \citenamefont {Ehlers}, \citenamefont {Tsurkan},\
  and\ \citenamefont {Loidl}}]{Kezsmarki2015}%
  \BibitemOpen
  \bibfield  {author} {\bibinfo {author} {\bibfnamefont {I.}~\bibnamefont
  {Kezsmarki}}, \bibinfo {author} {\bibfnamefont {S.}~\bibnamefont {Bordacs}},
  \bibinfo {author} {\bibfnamefont {P.}~\bibnamefont {Milde}}, \bibinfo
  {author} {\bibfnamefont {E.}~\bibnamefont {Neuber}}, \bibinfo {author}
  {\bibfnamefont {L.~M.}\ \bibnamefont {Eng}}, \bibinfo {author} {\bibfnamefont
  {J.~S.}\ \bibnamefont {White}}, \bibinfo {author} {\bibfnamefont {H.~M.}\
  \bibnamefont {Ronnow}}, \bibinfo {author} {\bibfnamefont {C.~D.}\
  \bibnamefont {Dewhurst}}, \bibinfo {author} {\bibfnamefont {M.}~\bibnamefont
  {Mochizuki}}, \bibinfo {author} {\bibfnamefont {K.}~\bibnamefont {Yanai}},
  \bibinfo {author} {\bibfnamefont {H.}~\bibnamefont {Nakamura}}, \bibinfo
  {author} {\bibfnamefont {D.}~\bibnamefont {Ehlers}}, \bibinfo {author}
  {\bibfnamefont {V.}~\bibnamefont {Tsurkan}}, \ and\ \bibinfo {author}
  {\bibfnamefont {A.}~\bibnamefont {Loidl}},\ }\href {\doibase
  10.1038/NMAT4402} {\bibfield  {journal} {\bibinfo  {journal} {Nature
  Materials}\ }\textbf {\bibinfo {volume} {14}},\ \bibinfo {pages} {1116}
  (\bibinfo {year} {2015})}\BibitemShut {NoStop}%
\bibitem [{\citenamefont {Khalyavin}\ \emph {et~al.}(2012)\citenamefont
  {Khalyavin}, \citenamefont {Manuel},\ and\ \citenamefont
  {Chapon}}]{Khalyavin2012}%
  \BibitemOpen
  \bibfield  {author} {\bibinfo {author} {\bibfnamefont {D.~D.}\ \bibnamefont
  {Khalyavin}}, \bibinfo {author} {\bibfnamefont {P.}~\bibnamefont {Manuel}}, \
  and\ \bibinfo {author} {\bibfnamefont {L.~C.}\ \bibnamefont {Chapon}},\
  }\href {\doibase 10.1103/PhysRevB.85.220401} {\bibfield  {journal} {\bibinfo
  {journal} {Physical Review B}\ }\textbf {\bibinfo {volume} {85}},\ \bibinfo
  {pages} {220401} (\bibinfo {year} {2012})}\BibitemShut {NoStop}%
\bibitem [{\citenamefont {Roessler}\ \emph {et~al.}(2006)\citenamefont
  {Roessler}, \citenamefont {Bogdanov},\ and\ \citenamefont
  {Pfleiderer}}]{Roessler2006}%
  \BibitemOpen
  \bibfield  {author} {\bibinfo {author} {\bibfnamefont {U.~K.}\ \bibnamefont
  {Roessler}}, \bibinfo {author} {\bibfnamefont {A.~N.}\ \bibnamefont
  {Bogdanov}}, \ and\ \bibinfo {author} {\bibfnamefont {C.}~\bibnamefont
  {Pfleiderer}},\ }\href {\doibase 10.1038/nature05056} {\bibfield  {journal}
  {\bibinfo  {journal} {Nature}\ }\textbf {\bibinfo {volume} {442}},\ \bibinfo
  {pages} {797} (\bibinfo {year} {2006})}\BibitemShut {NoStop}%
\bibitem [{\citenamefont {Butenko}\ \emph {et~al.}(2010)\citenamefont
  {Butenko}, \citenamefont {Leonov}, \citenamefont {Rossler},\ and\
  \citenamefont {Bogdanov}}]{Butenko2010}%
  \BibitemOpen
  \bibfield  {author} {\bibinfo {author} {\bibfnamefont {A.~B.}\ \bibnamefont
  {Butenko}}, \bibinfo {author} {\bibfnamefont {A.~A.}\ \bibnamefont {Leonov}},
  \bibinfo {author} {\bibfnamefont {U.~K.}\ \bibnamefont {Rossler}}, \ and\
  \bibinfo {author} {\bibfnamefont {A.~N.}\ \bibnamefont {Bogdanov}},\ }\href
  {\doibase 10.1103/PhysRevB.82.052403} {\bibfield  {journal} {\bibinfo
  {journal} {Physical Review B}\ }\textbf {\bibinfo {volume} {82}},\ \bibinfo
  {pages} {052403} (\bibinfo {year} {2010})}\BibitemShut {NoStop}%
\bibitem [{\citenamefont {Okubo}\ \emph {et~al.}(2012)\citenamefont {Okubo},
  \citenamefont {Chung},\ and\ \citenamefont {Kawamura}}]{Okubo2012}%
  \BibitemOpen
  \bibfield  {author} {\bibinfo {author} {\bibfnamefont {T.}~\bibnamefont
  {Okubo}}, \bibinfo {author} {\bibfnamefont {S.}~\bibnamefont {Chung}}, \ and\
  \bibinfo {author} {\bibfnamefont {H.}~\bibnamefont {Kawamura}},\ }\href
  {\doibase 10.1103/PhysRevLett.108.017206} {\bibfield  {journal} {\bibinfo
  {journal} {Physical Review Letters}\ }\textbf {\bibinfo {volume} {108}},\
  \bibinfo {pages} {017206} (\bibinfo {year} {2012})}\BibitemShut {NoStop}%
\end{thebibliography}
\end{document}
